%% file: arxiv.tex
\documentclass[a4paper,10pt]{article}
\usepackage{pdf14}
\usepackage[T1]{fontenc}
\usepackage[utf8]{inputenc}
\usepackage[l2tabu, orthodox]{nag}
\usepackage{geometry}
\usepackage[american]{babel}
\usepackage{caption}
\usepackage{subcaption}

\usepackage[tbtags]{amsmath}
\usepackage{amsthm,amssymb}
\usepackage{bm}

\usepackage{graphicx}
\usepackage{float}
\usepackage{wrapfig}

\usepackage{array}
\usepackage{booktabs}

\usepackage[hyphens]{url}
\usepackage[pdfa]{hyperref}
\hypersetup{
  colorlinks = true,
  allcolors = blue,
  pdfencoding = auto,
  psdextra
}

\usepackage{cleveref}
\usepackage{import}

\theoremstyle{plain}

\usepackage{stmaryrd}

\usepackage{colonequals}

\newenvironment{subsubfigure}[2][]{%
  \begin{subfigure}[#1]{#2}%
    \stepcounter{subsubfigure}%
}{%
    \addtocounter{subfigure}{-1}%
  \end{subfigure}%
}
\newcounter{subsubfigure}

\numberwithin{equation}{section}

\begin{document}

\title{A spectrum of p-atic symmetries and defects in confluent epithelia}

\author{Lea Happel, Griseldis Oberschelp, Anneli Richter, Gwenda Roselene\\ Rode, Valeriia Grudtsyna, Amin Doostmohammadi, Axel Voigt}

\maketitle

\input{macros_Defects}

\begin{abstract}
Topological defects provide a unifying language to describe how orientational order breaks down in active and living matter. Considering cells as elongated particles confluent, epithelial tissues can be interpreted as nematic fields and its defects have been linked to extrusion, migration, and morphogenetic transformations. Yet, epithelial cells are not restricted to nematic order: their irregular shapes can express higher rotational symmetries, giving rise to $p$-atic order with $p>2$. Here we introduce a framework to extract $p$-atic fields and their defects directly from experimental images. Applying this method to MDCK cells, we find that all symmetries from $p=2$ to $p=6$ generate $\pm \frac{1}{p}$ defects. Surprisingly, the statistics reveal an even–odd asymmetry, with odd $p$ producing more defects than even $p$, consistent with geometric frustration arguments based on tilings. In contrast, no strong positional or orientational correlations are found between nematic and hexatic defects, suggesting that different symmetries coexist largely independently. These results demonstrate that epithelial tissues should not be described by nematic order alone, but instead host a spectrum of $p$-atic symmetries. Our work provides the first direct experimental evidence for this multivalency of order and offers a route to test and refine emerging $p$-atic liquid crystal theories of living matter.
\end{abstract}

\section*{Introduction}
Topological defects—localized disruptions of orientational order—are emerging as universal signatures of organization in living matter. In confluent epithelia, where cell shapes define coarse-grained orientational fields, the creation, motion, and annihilation of defects are closely tied to morphogenetic processes. There is increasing evidence that the shape of cells and resulting orientational order and associated topological defects are essential for mechanical mechanisms that drive morphogenetic processes during embryonic development. Considering for example the elongation of cells to define a coarse-grained nematic order \cite{duclos2017topological,saw2017topological,kawaguchi2017topological} allows to describe properties of the tissue. Corresponding topological defects have been related to cell extrusions \cite{saw2017topological,monfared2023mechanical}, morphological changes \cite{maroudas2021topological} and active turbulence \cite{alert2022active}. The underlying coarse-grained theories are nematic liquid crystal theories based on Oseen–Frank and Landau–de Gennes energies \cite{de_Gennes_book}, which if combined with flow lead to Ericksen–Leslie and Beris–Edwards models, respectively. These theories have successfully captured a wide range of epithelial phenomena and established defects as key organizing elements of tissue mechanics.

\begin{figure*}[!hb]
    \vspace{1em}
    \centering
    \includegraphics[width=\linewidth]{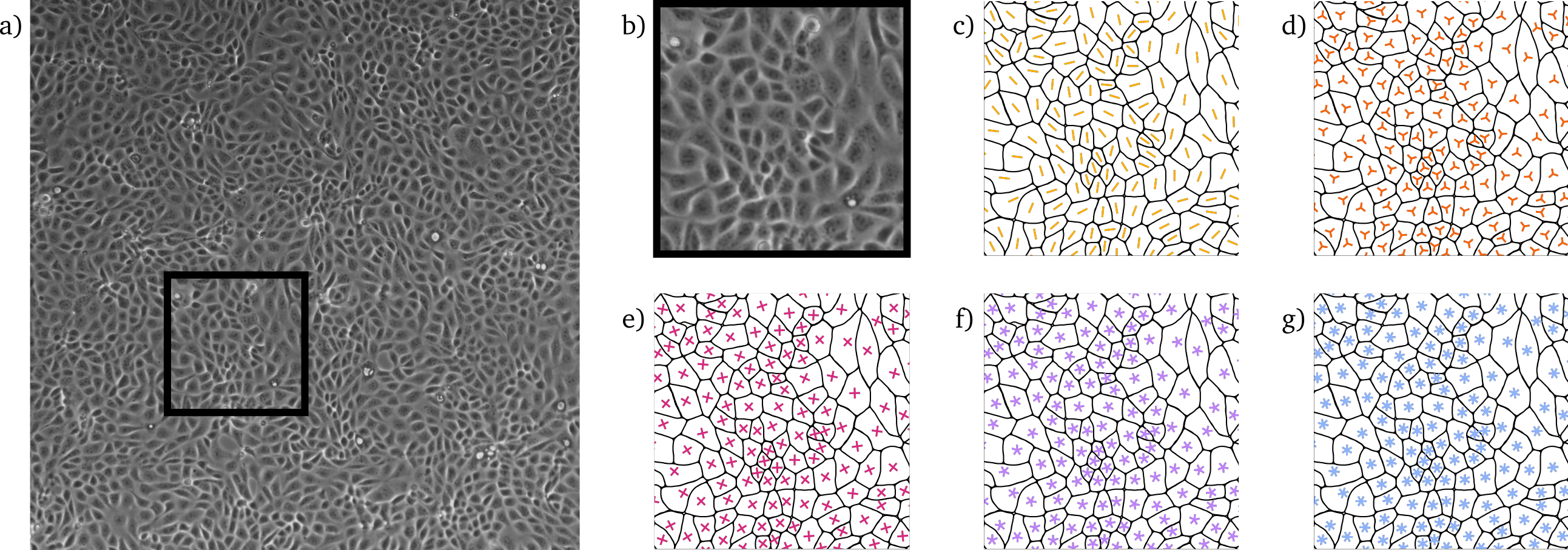}
    \caption{{\bf Shape classification of cells in wild-type MDCK cell monolayer.} a) Raw experimental data - Full data of Frame 0. b) Raw experimental data - used snippet of Frame 0 for visualization c) - g) Minkowski tensor for cells in b), visualized using $\vartheta_p$ for $p = 2,3,4,5,6$, respectively. The rotation of the $p$-atic director indicates the orientation.}
    \label{fig:2_Minkowski}
\end{figure*}

Beyond the nematic case ($p=2$), higher order rotational symmetries have recently been reported in both biological and synthetic systems, e.g. \cite{Cislo_NaturePhysics_2023} for tetratic order ($p=4$) and \cite{li2018role,durand2019thermally,pasupalak2020hexatic,armengol2023epithelia,eckert2023hexanematic} for hexatic ($p=6$) order. In contrast with nematic order, which represents rotational symmetries under rotation by $\pi$, tetratic and hexatic order considers rotational symmetry under rotation by $\frac{\pi}{2}$ and $\frac{\pi}{3}$, respectively. This classification can be extended to define $p$-atic order, which considers rotational symmetry under rotation by $\frac{2 \pi}{p}$, with $p$ an integer. Liquid crystal phases with $p$-atic order have a long history, starting with Onsager who showed that the symmetry and shape of constituent particles can lead to $p$-atic order \cite{j.1749-6632.1949.tb27296.x}. Most prominently, hexatic order has been postulated and found by experiments and simulations as an intermediate state between crystalline solid and isotropic liquid in \cite{PhysRevLett.41.121,PhysRevB.19.2457,PhysRevLett.58.1200,PhysRevLett.74.2519,PhysRevLett.82.2721,cphc.200900755,PhysRevLett.107.155704}. Other examples are colloidal systems for triadic platelets \cite{Bowick_2017} or cubes \cite{Wojciechowski}, which lead to $p$-atic order with $p=3$ and $p=4$, respectively. Even pentatic ($p=5$) and heptatic ($p=7$) liquid crystals have been engineered by combining different colloidal tiles \cite{wang2018brownian,YU2023}. First models for such systems go back to \cite{Nelson_JP_1987} and can be viewed as extensions of Oseen–Frank-like models, accounting for higher rotational symmetries of the director field. More recently liquid crystal theories for $p$-atic order which extend the Landau–de Gennes approach have been proposed \cite{giomi2022hydrodynamic,giomi2022long}. These models consider higher order $Q$-tensors and couple them with fluid flow. Active extensions of these theories predict defect creation, currents, and instabilities whose quantization depends on $p$, pointing to qualitatively new physics in living matter. These theories have already been used to model epithelial tissue \cite{krommydas2023hydrodynamic,armengol2023epithelia,armengol2024hydrodynamics}. What remains open, however, is how the irregular and dynamic shapes of epithelial cells give rise to $p$-atic fields and defects, and how these connect to the continuum predictions. Our work addresses this gap by showing that epithelial cell shapes naturally encode multiple $p$-atic orders and by directly quantifying their defect content.

Rather than proposing a new methodology per se, we use Minkowski tensor–based shape descriptors as a lens to reveal previously hidden physics: the coexistence of distinct $p$-atic orders and their defect networks in confluent epithelial tissues. Our approach starts from quantified rotational symmetries of rounded and irregular cell shapes, see Figure \ref{fig:2_Minkowski}. These data show a snapshot of the raw experimental data of a MDCK (Madin–Darby Canine Kidney) cell monolayer, together with segmented cell shapes and their shape classification by Minkowski tensors for various $p$, visualized by the orientation $\vartheta_p$, see \cite{Happel2025.01.03.631196} for details. Minkowski tensors have been shown to be robust and advantageous to other shape characterization methods, such as the bond order parameter \cite{Loewe_PRL_2020, monfared2023mechanical} or the shape function \cite{armengol2023epithelia}. Using these per-cell measurements as inputs, we construct continuous $p$-atic fields at user-controlled coarse-graining radii. We provide visualization tools that help to identify the essential information of $p$-atic liquid crystals. Here we draw connections to computer graphics, where director fields with higher rotational symmetries are known as rotational symmetry (RoSy) fields \cite{10.1145/1356682.1356683,10.1145/1276377.1276446} and also the connection between higher order $Q$-tensor fields and these RoSy-fields has been established and used for visualization \cite{10.1145/1276377.1276446}. We essentially follow \cite{Palacios_IEEE_2011} and use the line integral convolution (LIC) technique for each direction and blend the results. This provides a high contrast texture-based image of the coarse-grained orientation $\vartheta_p$. The second aspect concerns the identification of topological defects. While various approaches for this task have been proposed \cite{wenzel2021defects,skogvoll2023unified} we here follow our experience in nematic fields and consider the intersection points of zero contour lines to identify defects. This provides the points at which the $Q$-tensor is singular. As also for higher order $Q$-tensors the irreducible information is contained in only two linearly independent components, their zero-contours are sufficient to determine the defects. After the locations are identified the topological charge follows by computing the winding number. 

We only found defects of topological charge $\pm \frac{1}{p}$. While other defects are theoretically possible the ones found are energetically most favorable. Placing these defects atop the LIC textures provides an intuitive, symmetry-aware overview that scales cleanly from $p=2$ to higher $p$. Statistics on these defects reveal an even-odd asymmetry but do not show strong correlations between defects of different $p$-atic order. Together, these analyses uncover that epithelial tissues harbor a spectrum of topological defects across multiple symmetries—an observation that calls for new coarse-grained descriptions beyond the nematic paradigm.

The provided tools are not only applicable for experimental data of confluent monolayers of MDCK cells, the tools can also be applied to computational results of cell-resolved models \cite{alert2020physical}, as well as coarse-grained models for $p$-atic liquid crystals \cite{giomi2022hydrodynamic,giomi2022long} and therefore allow for a direct comparison. By establishing this bridge, we enable direct confrontation between cell-resolved data and continuum $p$-atic theories, and reveal new physics in how multiple orientational symmetries coexist and organize in living tissues. For all tasks we provide the considered Python code, see \cite{zenodo}.

\section*{Materials and methods}
All steps from cell shapes to tissue-scale $p$-atic defects are illustrated in Figure \ref{fig:Explanation_Defects}. Therefore again the snippet from Figure \ref{fig:2_Minkowski} $b)$ is used. We briefly motivate each step physically before giving the implementation.

\subsection*{Computing (irreducible) Minkowski tensors for each cell}

The first step for constructing a global $p$-atic field is to calculate $\vartheta_p$ and $q_p$ for every cell. As the general framework used for this - (irreducible) Minkowski tensors or equivalently higher order Q-tensors - is already described in detail in numerous publications, we will only motivate the method and focus on the specific data format. Physically, these descriptors quantify how closely a cell shape expresses a $p$-fold rotational symmetry (e.g., rod-like for $p{=}2$, square-like for $p{=}4$, hexagon-like for $p{=}6$), and thus generalize nematic order to arbitrary $p$.

From the microscopy image, which can be seen for example in Figure \ref{fig:2_Minkowski} $a)$, a cell segmentation is generated. This segmentation is stored as grayscale image, whereby every cell is associated to a different value. Firstly we extract the contour $\mathcal{C}$ of each cell, using the \textit{contour} function from \textit{scikit-image} \cite{scikit-image}. To remove pixel-shaped artifacts, we slightly smooth out this contour $\mathcal{C}$. This smoothing avoids spurious high-frequency contributions to $\psi_p$ from jagged pixel edges and yields robust shape integrals. We can then use the outward pointing normal $\mathbf{n}$ of this contour to calculate  
\begin{align}
    \psi_p({\cal{C}})=\frac{1}{2\pi}\int_{\partial\cal{C}} e^{ip\theta_{\boldsymbol{n}}}\,{\rm d} \partial\mathcal{C},
\end{align}
with $\theta_{\boldsymbol{n}}$ the orientation of the outward pointing normal $\mathbf{n}$.  The complex phase of $\psi_p$ encodes the preferred $p$-atic orientation
\begin{align}
\vartheta_p({\cal{C}})=\frac{1}{p} \arctan2 \left(\Im \psi_p({\cal{C}}),\Re \psi_p({\cal{C}}) \right) + \frac{\pi}{p},
\end{align}
with $\Im \psi_p$ and $\Re \psi_p$ the imaginary and real part of $\psi_p$, respectively, whereas its magnitude 
\begin{align}
q_p({\cal{C}}) = \frac{|\psi_p({\cal{C}})|}{\psi_0({\cal{C}})},
\end{align}
with $\psi_0({\cal{C}})$ corresponding to the length of the contour multiplied with the factor $\frac{1}{2\pi}$, measures how strongly the shape exhibits $p$-fold symmetry. The division by $\psi_0({\cal{C}})$ is needed to ensure scale invariance. In other words, $q_p$ is a dimensionless “strength” of $p$-aticity independent of cell size.

\begin{figure*}[pbht]
    \centering
    \includegraphics[width=\linewidth]{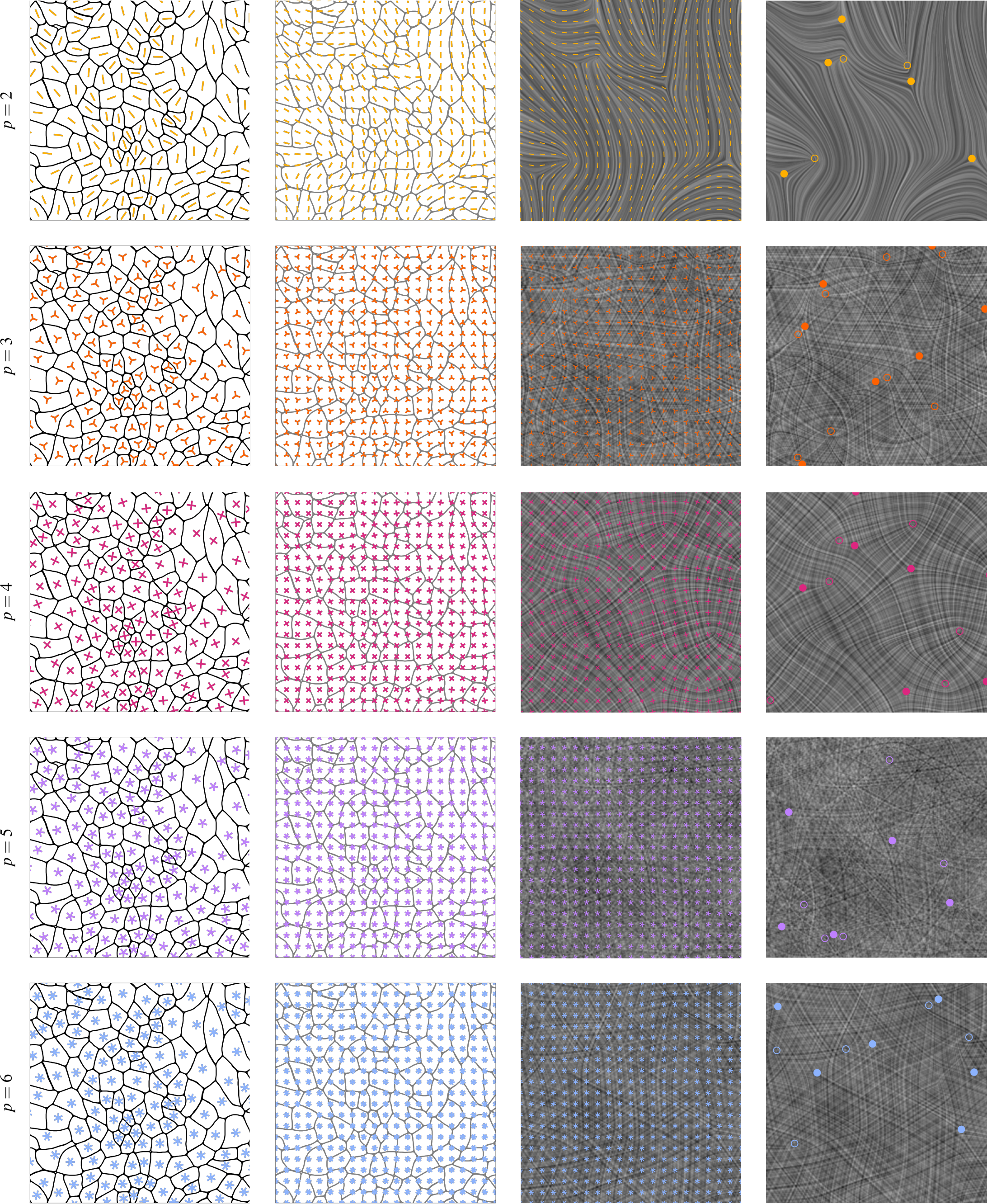}
    \caption{Illustration of how we go from the orientation of cells (left column) to a fine orientation field (middle columns), which we then visualize as a LIC image (the two right  most columns) and use to detect defects (right column). In this illustration a coarse-graining radius $\ravg$ of $1.5r_{max}(t)$ is used. Positively charged defects are indicated by an open circle and negatively charged defects by a closed circle. Different colors are used for different $p$, corresponding to the color scheme used in Figure \ref{fig:2_Minkowski}. The pictures on the left correspond to the pictures on Figure \ref{fig:2_Minkowski}.}
    \label{fig:Explanation_Defects}
\end{figure*}

In this paper we are mainly concerned with the orientation $\vartheta_p$. An illustration of the orientation of different cells for ${p \in [2,3,4,5,6]}$ is shown in Figures \ref{fig:2_Minkowski} and \ref{fig:Explanation_Defects}. These per-cell orientations are the inputs to the tissue-level fields analyzed below.

\subsection*{Generation of coarse-grained $p$-atic fields}
We now use the orientations $\vartheta_p$ of the individual cells to obtain a global orientation field via interpolation. As we understand the global orientation field and with this its defects as a tissue and not a cell property we also use some averaging in the process to smooth the global field. Of course the extent of this averaging will influence the obtained field and with it the number and the location of the defects. Roughly said more averaging leads to smoother fields and a smaller number of defects. To account for this we consider three different radii for averaging to investigate if the qualitative behavior is independent of the extent of averaging. We will also refer to this averaging as coarse-graining in the following. Physically, the coarse-graining radius $\ravg$ defines the observation scale: larger $\ravg$ probes longer wavelengths and emphasizes tissue-level organization over cell-scale variability.

For the context of averaging, interpolating and finally detecting the defect location we will not use $\vartheta_p$ and $q_p$ but the related higher order Q-tensor representation. The reason for this is twofold: On the one hand, Q-tensors of order $p$ naturally encode the invariance under rotations of $\frac{2\pi}{p}$. On the other hand these Q-tensor fields are well suited for defect detection, as defects are singular points in this defect field, meaning points where all linear independent components of this Q-tensor are zero. This representation also avoids branch-cut ambiguities in angles and makes zero-set geometry well defined.

In $2D$ Q-tensors have effectively two independent components $\QTensorCompA$ and $\QTensorCompB$, for the reduced (irreducible) description we use, independent of order $p$. We therefore firstly calculate these two components from $\vartheta_p(\mathcal{C})$ as \begin{align}
    \QTensorCompA(\mathcal{C})=\cos(p\vartheta_p(\mathcal{C})), \qquad \QTensorCompB(\mathcal{C})=\sin(p\vartheta_p(\mathcal{C})).
\end{align}
Thereby all elements $Q(\mathcal{C})_{i_1i_2...i_p}$ of $Q(\mathcal{C})$ with an even number of ones in the indexes are proportional to $ \QTensorCompA(\mathcal{C})$ and all elements with an odd number of ones in the indexes are proportional to $\QTensorCompB(\mathcal{C})$ \cite{armengol2023epithelia}.
Note that in the implementation summation formulas for this are used. Expressing the field as $(\QTensorCompA,\QTensorCompB)$ means that all physically equivalent orientations related by $2\pi/p$ map to the same point on the unit circle in this space.

We then calculated an averaged and interpolated orientation for each point on the fine regular grid to obtain a global field. We denote the $j$-th cell as $\mathcal{C}_j$ and its midpoint as $c_j$. Then the formula for the value of a grid point $x_{fine}^{i}$ on the fine grid reads as follows:
\begin{align}
     \QTensorCompAInt(x_{fine}^{i})=\sum_{j} w^{int}_{\mathcal{C}_j}(x_{fine}^{i})\frac{\QTensorCompA(\mathcal{C}_j)}{\sqrt{{\QTensorCompA (\mathcal{C}_j)}^2+{\QTensorCompB(\mathcal{C}_j)}^2}} , \\
    \QTensorCompBInt(x_{fine}^{i})=\sum_{j} w^{int}_{\mathcal{C}_j}(x_{fine}^{i})\frac{\QTensorCompB(\mathcal{C}_j)}{\sqrt{{\QTensorCompA (\mathcal{C}_j)}^2+{\QTensorCompB(\mathcal{C}_j)}^2}} .
\end{align}
Thereby we calculate the weights $w^{int}_{\mathcal{C}_j}$ as 
\begin{align}
w^{int}_{\mathcal{C}_j}(x_{fine}^{i})=\frac{g(||x_{fine}^{i}-c_j||_2)}{\sum_{k}g(||x_{fine}^{i}-c_k||_2)}
\end{align}
with 
\begin{align}
    g(d)=\max(\ravg-d,0.0)
\end{align}
a linear interpolation. This triangular kernel gives local, compact support and a transparent control of the averaging length. A larger value of $\ravg$ thereby means more interpolation and with this a smoother global field and fewer defects. We choose $\ravg$ in dependence on the largest cell radius $r_{max}(t)$ at the given time, whereby the cell radius here is understood as the largest distance of a point of the cell outline to the midpoint of the cell. In this paper we consider $\ravg=1.5r_{max}(t)$, $\ravg=2.25r_{max}(t)$ and $\ravg=3.0r_{max}(t)$. Results for all three values demonstrate that our qualitative conclusions are robust to the observation scale.

\subsection*{Localization of defects}
Defects are singularities in the Q-tensor field, which means the Q-tensor is the $0$-tensor at defects. As there are only two independent components this corresponds to $\QTensorCompAInt(x_{fine}^{i})=0.0$ and $\QTensorCompBInt(x_{fine}^{i})=0.0$. Geometrically, these are intersection points of the two zero-contours, i.e., locations where the local orientation is undefined. Exploiting this, we detect defect locations as the intersections of the contour lines of $\QTensorCompAInt(x_{fine})=0.0$ and $\QTensorCompBInt(x_{fine})=0.0$.

Other methods, precisely the usage of the Dirichlet energy or the Jacobian determinant, were tested, but in our data lead to a higher number of false-positive results. We therefore adopted the zero-level–set intersection criterion for its robustness and interpretability.

\subsection*{Calculation of the winding number}
To detect the type of defect we closely follow the definition of the winding number 
\begin{align}
    \omega = \frac{1}{2\pi} \oint_C d\vartheta.
\end{align}
Hereby $C$ denotes a closed path around the defect and $\vartheta$ the angle of the eigenvectors enclosed by the $x-$axis. For a $p$-atic field the topological charge is quantized in units of $\pm 1/p$, reflecting the $2\pi/p$ symmetry.

Implementation-wise, we extract a closed path around the defect. To be precise we use a square-path for this, where the sides of the square have a distance of $4$ pixels to the defect location. We then extract the corresponding eigenvectors along this path and align their orientation. Aligning the orientation means that we start at an arbitrary point on the path with an arbitrary orientation. Arbitrary here refers to the fact that we have $p$ equivalent directions to choose from, which correspond to the $p$ legs of the $p$-atic star. The resulting winding number is independent of the choice of the starting orientation, therefore this is arbitrary. We then consider two neighboring orientations, step by step in counter-clockwise direction. Aligning now means that we always consider the two orientations which enclose an angle $\leq \frac{\pi}{p}$. Note that it is crucial to close the path by aligning the last eigenvector of the path with a copy of the eigenvector with which we started. Summing up all the angles between two neighboring vectors and dividing this sum by $2\pi$ gives the winding number. This method also ensures that the resulting winding number is a multiple of $\pm \frac1p$. This alignment procedure removes artificial jumps and ensures that the measured $\omega$ captures the intrinsic rotation of the field.

Note that this method assumes that there are no jumps of $\frac \pi p$ or larger between two neighboring vectors. In practice, this assumption is satisfied away from regions of very low $q_p$, where orientations are poorly defined.
\subsection*{Calculation of the direction of \texorpdfstring{$+\frac{1}{2}$}{+1/2} defects}
\begin{figure*}[!hbt]
    \centering
    \includegraphics[width=\linewidth]{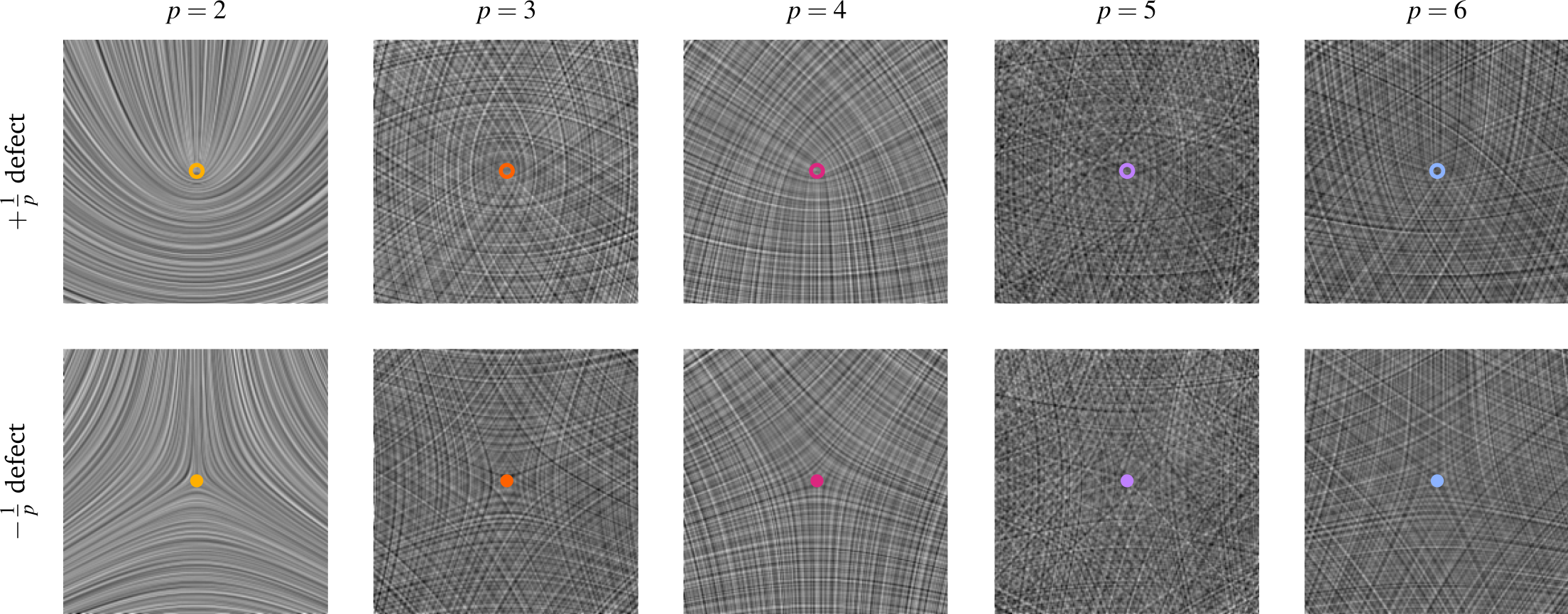}
    \caption{$+\frac{1}{p}$ and $-\frac{1}{p}$ defects shown as described in \cite{Palacios_IEEE_2011}. For the LIC filter the \textit{vtkSurfaceLICMapper} from \textit{vtk} \cite{vtkBook} was used, which is implemented based on the ideas from \cite{Laramee_IEEE_2003}. Positively charged defects are indicated by an open circle and negatively charged defects by a closed circle. Different colors are used for different $p$, corresponding to the color scheme used in Figure \ref{fig:2_Minkowski}.}
    \label{fig:Perfect_Defects}
\end{figure*}

Defects with a winding number of $+\frac{1}{2}$ are comet-shaped, as can be seen for example in Figure \ref{fig:Perfect_Defects}, and with this have a directionality. This polarity is unique among defects in our dataset and enables spatio-orientational comparisons across symmetries. To calculate this we follow \cite{Vromans_SM_2016, Wenzel_PRE_2021}. We calculate a tensor 
\begin{align*}
 \mathbf{N^{i}}=
 \begin{pmatrix}
     \cos{\vartheta_2^{i}}\cos{\vartheta_2^{i}} & \cos{\vartheta_2^{i}}\sin{\vartheta_2^{i}} \\
     \cos{\vartheta_2^{i}}\sin{\vartheta_2^{i}} & \sin{\vartheta_2^{i}}\sin{\vartheta_2^{i}}
 \end{pmatrix}   
\end{align*}
whereby $\vartheta_2^{i}$ denotes the orientation for $p=2$ at the grid-point $i$ of the fine regular grid. Then we get the orientation $\vartheta_{+\frac12}$ of the defect as
\begin{align*}
   \vartheta_{+\frac12}=\arctan2(\langle \partial_x N^{i}_{xy} + \partial_y N^{i}_{yy}\rangle, \langle \partial_x N^{i}_{xx} + \partial_y N^{i}_{xy}\rangle ).
\end{align*}
Hereby $\langle \cdot,\cdot \rangle$ denotes the average along a short loop around the $+\frac12$ in question. For this loop we again choose a square-path, where the sides of the square have a distance of $4$ pixels to the defect location. We use this direction in Section \nameref{sec:Results} to test for alignment with hexatic defects.
\subsection*{Visualization}
To visualize the $p$-atic fields we follow \cite{Palacios_IEEE_2011} and use the line integral convolution (LIC) technique for each direction and blend the resulting images. 
This provides a high contrast texture-based image of the coarse-grained orientation $\vartheta_p$. For higher $p$, the blending across equivalent directions yields symmetry-aware textures in which the $p$-leg structure is visually apparent. Topological defects already become visible as singular points in these fields. To further guide the eye we plot the identified defects together with their charge on top. Figure \ref{fig:Perfect_Defects} provides a classification of the considered $\pm \frac{1}{p}$ defects with $p=2,3,4,5,6$.

\subsection*{Experimental setup}
\textbf{Cell culture}
Madin-Darby canine kidney (MDCK) cells were cultured in DMEM (DMEM, low glucose, GlutaMAX$^{TM}$ Supplement, pyruvate) supplemented with 10$\%$ fetal bovine serum (FBS; Gibco) and 100 $U/mL$ penicillin/streptomycin (Gibco) at 37$^\circ$ C with 5$\%$ $CO_2$. The cell line was tested for mycoplasma. MDCK monolayers provide reproducible epithelial organization and robust junctions, which are advantageous for quantifying tissue-scale orientational order and defects.

\noindent
\textbf{Monolayer preparation}
Cells were seeded on glass-bottom dishes (Mattek) pretreated with 10 $\mu g/mL$ fibronectin human plasma in phosphate buffered saline (PBS, pH 7.4; Gibco). Fibronectin was incubated for 30 min at 37$^\circ$ C. The initial cell seeding density was sparse. They were imaged approximately 24 hours after, when a confluent monolayer was formed. Uniform ECM coating promotes consistent adhesion and spreading, so that cell shapes evolve primarily due to crowding and neighbor interactions—key drivers of $p$-atic order in epithelia.

\noindent
\textbf{Live cell imaging}
Samples were imaged using Nikon ECLIPSE Ti microscope equipped with a H201-K-FRAME Okolab chamber, heating system (Okolab) and a 
$CO_2$ pump (Okolab) which maintained them at 37$^\circ$ C and at 5$\%$ $CO_2$. Phase microscopy images were taken every 10 minutes using a 10$\times$ NA=0.3 Plan Fluor objective and Andor Neo 5.5 sCMOS camera. This cadence resolves cell-shape fluctuations and collective rearrangements on the timescales relevant for defect creation, motion, and annihilation, while minimizing phototoxicity by avoiding fluorescence.

\noindent
\textbf{Image analysis}
The time-series were xy-drift corrected using Fast4DReg plugin \cite{Laine_2019}, \cite{Pylv504744} in FIJI. Cells were segmented and tracked using the python module CellSegmentationTracker (\url{https://github.com/simonguld/CellSegmentationTracker}), which utilizes both Cellpose \cite{cellpose2021} and Trackmate \cite{trackmate2017}. A Cellpose model was trained by manual segmentation of the phase contrast images. Accurate contours are essential because they are the sole inputs to the Minkowski-tensor integrals that determine $\vartheta_p$ and $q_p$; we therefore trained a model on our imaging conditions to maximize segmentation fidelity.

\section*{Results}
\label{sec:Results}
\begin{figure*}[!htb]
    \centering
    \includegraphics[width=0.9\linewidth]{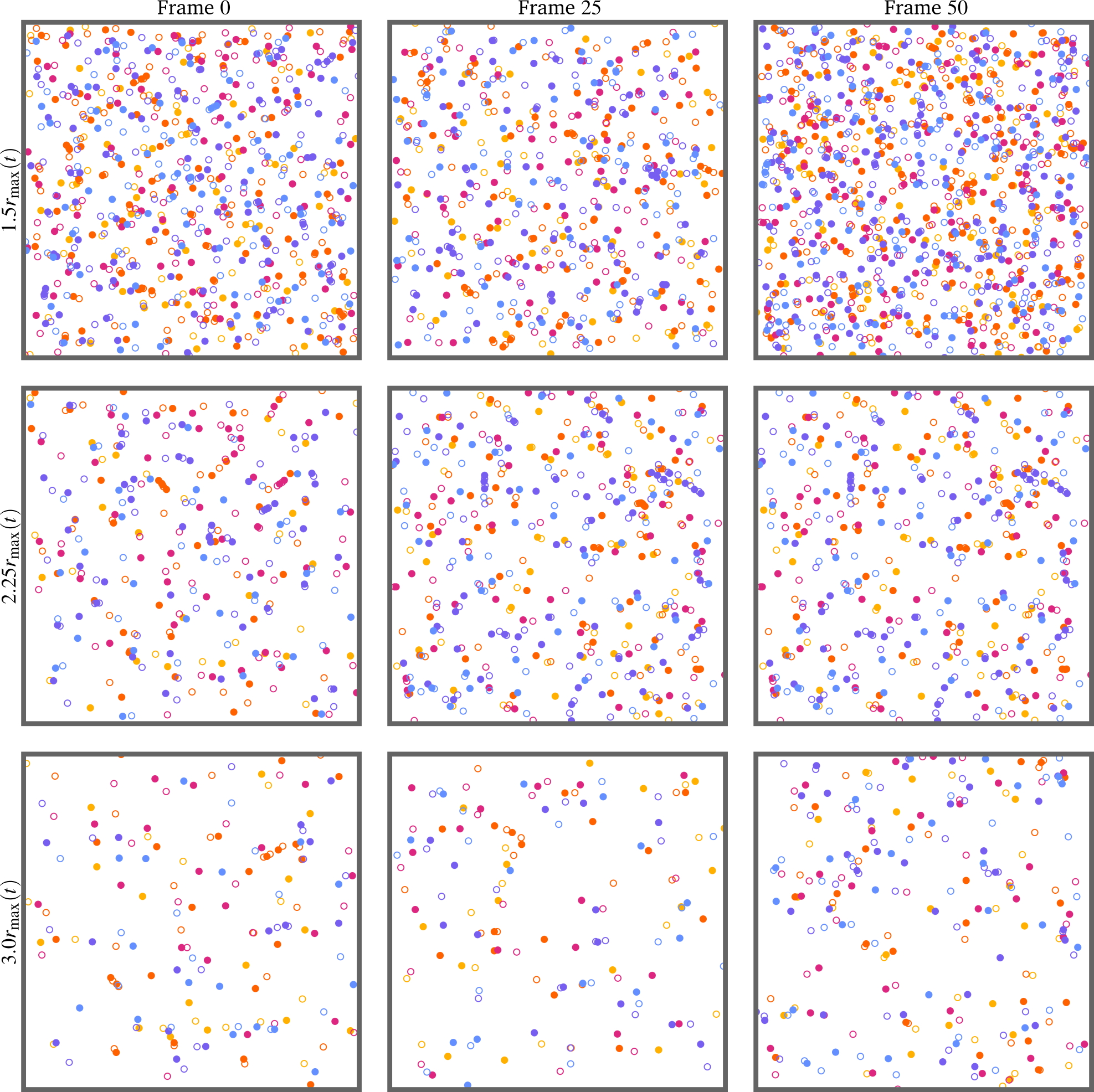}
    \caption{Defect position for different coarse-graining radii $\ravg$ at different times. Different colors are used for different $p$, corresponding to the color scheme used in Figure \ref{fig:2_Minkowski}. Positively charged defects are indicated by an open circle and negatively charged defects by a closed circle. The raw experimental image for Frame 0 can be seen in Figure \ref{fig:2_Minkowski} $a)$, the one for Frame 25 in Figure \ref{fig:frame_25} and the one for Frame 50 in Figure \ref{fig:frame_50}.}
    \label{fig:scatter_plot}
\end{figure*}
We are mainly interested in qualitative correlations between defects of different orders. Before turning to such correlations, we first establish the basic statistics of defect formation across scales of observation. As there is no clear indication about the best coarse-graining radius $\ravg$ we systematically explore three different radii, $\ravg=1.5r_{max}(t)$, $\ravg=2.25r_{max}(t)$ and $\ravg=3.0r_{max}(t)$. We use the full frames of the experimental data to calculate a global orientation field and the $p$-atic defects thereof. For all three different radii only $\pm\frac{1}{p}$ defects were detected. This observation is fully consistent with continuum theory, where $\pm \frac{1}{p}$ charges are the lowest-energy topological excitations in $p$-atic fields.

As can be already seen in Figure \ref{fig:scatter_plot} a bigger coarse graining radius leads to a smaller number of defects. This scale dependence reflects the fact that coarse-graining suppresses short-wavelength fluctuations, smoothing out smaller defect– antidefect pairs. One can also already suspect from this, especially comparing frame $0$ and frame $50$, that the number of defects grows over time. This temporal increase suggests that as the monolayer becomes denser and more crowded, cell-shape irregularities continuously seed new orientational defects.

\begin{figure*}[!htb]
    \centering
    \includegraphics[width=\linewidth]{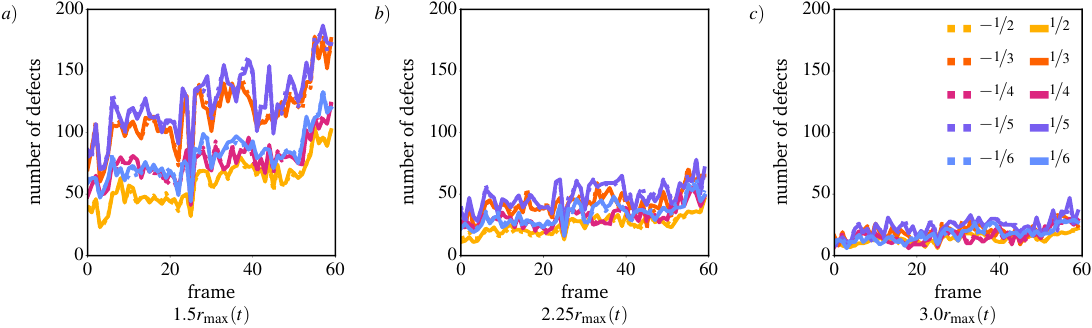}
    \caption{Number of $-\frac1p$ and $+\frac1p$ defects over time  for different coarse-graining radii $\ravg$. Different colors are used for different $p$, corresponding to the color scheme used in Figure \ref{fig:2_Minkowski}.}
    \label{fig:defects_over_time}
\end{figure*}

To quantify these observations we evaluate the number of $\pm \frac{1}{p}$ defects over time. For this and all following evaluations we exclude all defects with a distance smaller than $\ravg$ to any domain boundary to ensure that our evaluations do not include boundary effects. As can be seen in Figure \ref{fig:defects_over_time} for all defect types and all $\ravg$ the number of defects grows over time. This is likely connected to the growing number of cells and the growing cell density over time. As expected from topological charge conservation, the numbers of $+\frac1p$ and $-\frac1p$ defects are very close to each other and are barely distinguishable in Figure \ref{fig:defects_over_time}. Especially for $\ravg=1.5r_{max}(t)$ the numbers of defects corresponding to an odd $p$, so to $p=3$ or $p=5$, are noticeable higher than the numbers of defects for even $p$. This even–odd asymmetry becomes less prominent as $\ravg$ increases, suggesting that it originates from local tiling constraints of individual cell shapes rather than global tissue organization.

\begin{figure*}[!tbh]
    \centering
\includegraphics[width=0.8\linewidth]{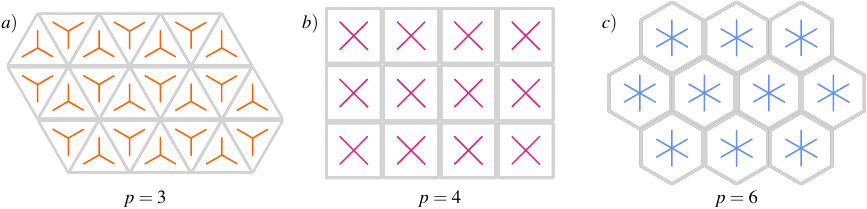}
    \caption{Tilings of the reference shapes for $p=3$ ($a)$ with equilateral triangles), $p=4$ ($b)$ with squares) and $p=6$ ($c)$ with hexagons).}
    \label{fig:explanation_reference_shape}
\end{figure*}

While it is hardly possible to pin down an exact reason for this, an intuitive explanation can be found by thinking about tilings made out of the corresponding reference shapes for different $p$. For $p=4$ and $p=6$ one can easily imagine a tiling of the space with only squares or only regular hexagons, see Figure \ref{fig:explanation_reference_shape} $b)$ and $c)$ for an illustration. Importantly, the orientation $\vartheta_4$ of the squares or $\vartheta_6$ of the hexagons is the same for all tiles. So it is possible to construct an ordered phase without any defects out of these reference shapes for $p=4$ and $p=6$. 

For $p=3$ and $p=5$ the situation is different. The reference shape corresponding to $p=3$ is an equilateral triangle. While it is possible to tile the space with this shape, neighboring triangles need to be rotated by $60^\circ$, see the illustration in Figure \ref{fig:explanation_reference_shape} $a)$. Therefore the associated orientations $\vartheta_3$ of neighboring triangles are orthogonal (with respect to $p=3$) to each other, and this tiling of the space is not ordered and therefore would also not be defect free. For $p=5$ the situation is even worse, as it is not possible to tile the space with regular pentagons. $p=2$ is not regarded, as the associated shape for this is a line and therefore degenerate. However, it is possible to construct an ordered tiling of the space with rectangles elongated along one axis, leading to a defect-free nematic state described by $\vartheta_2$. In these cases of tilings with the reference shapes it is natural to expect that the number of defects for odd $p$ will be higher than the number of defects for even $p$. Even though real epithelial tissues are far from perfect tilings and consist of highly irregular cell shapes, this geometric frustration argument provides a simple physical rationale for the observed odd–even asymmetry in defect statistics.

\begin{figure*}[tbh]
    \centering
    \includegraphics[width=\linewidth]{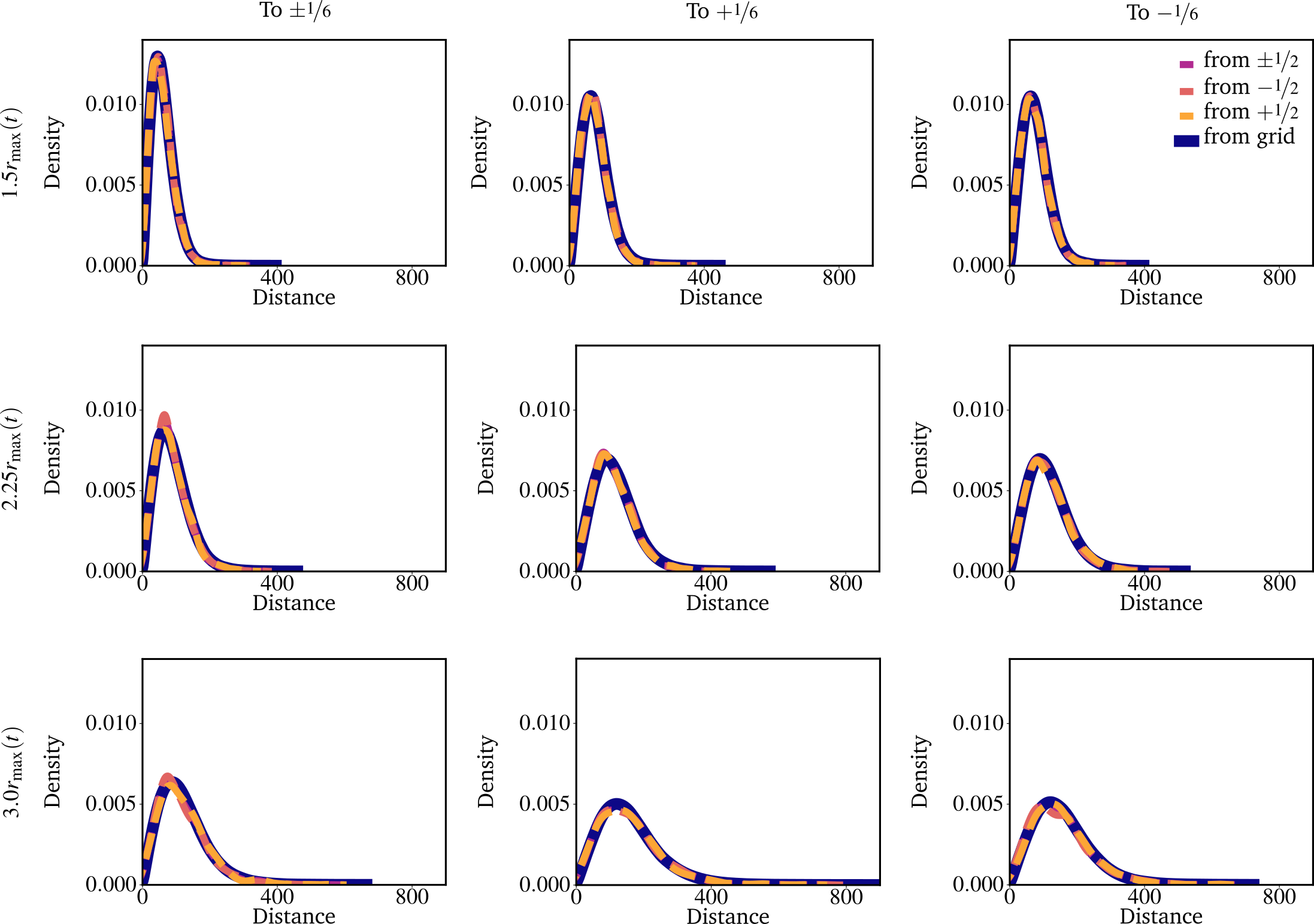}
    \caption{Kde plots of the density functions of the distance (measured in pixels) from $\pm \frac12$ ($-\frac12$, $+\frac12$) defects to the closest $\pm \frac16$ ($+\frac16$, $-\frac16$) defects and kde plots of the density functions of the distance from any gridpoint to the the closest $\pm \frac16$ ($+\frac16$, $-\frac16$) defects are shown for different coarse grain radii $\ravg$ (each row corresponds to one radius). The axis scaling is the same in all plots.}
    \label{fig:From_2_To_6}
\end{figure*}
\begin{figure*}[tbh]
    \centering
    \includegraphics[width=\linewidth]{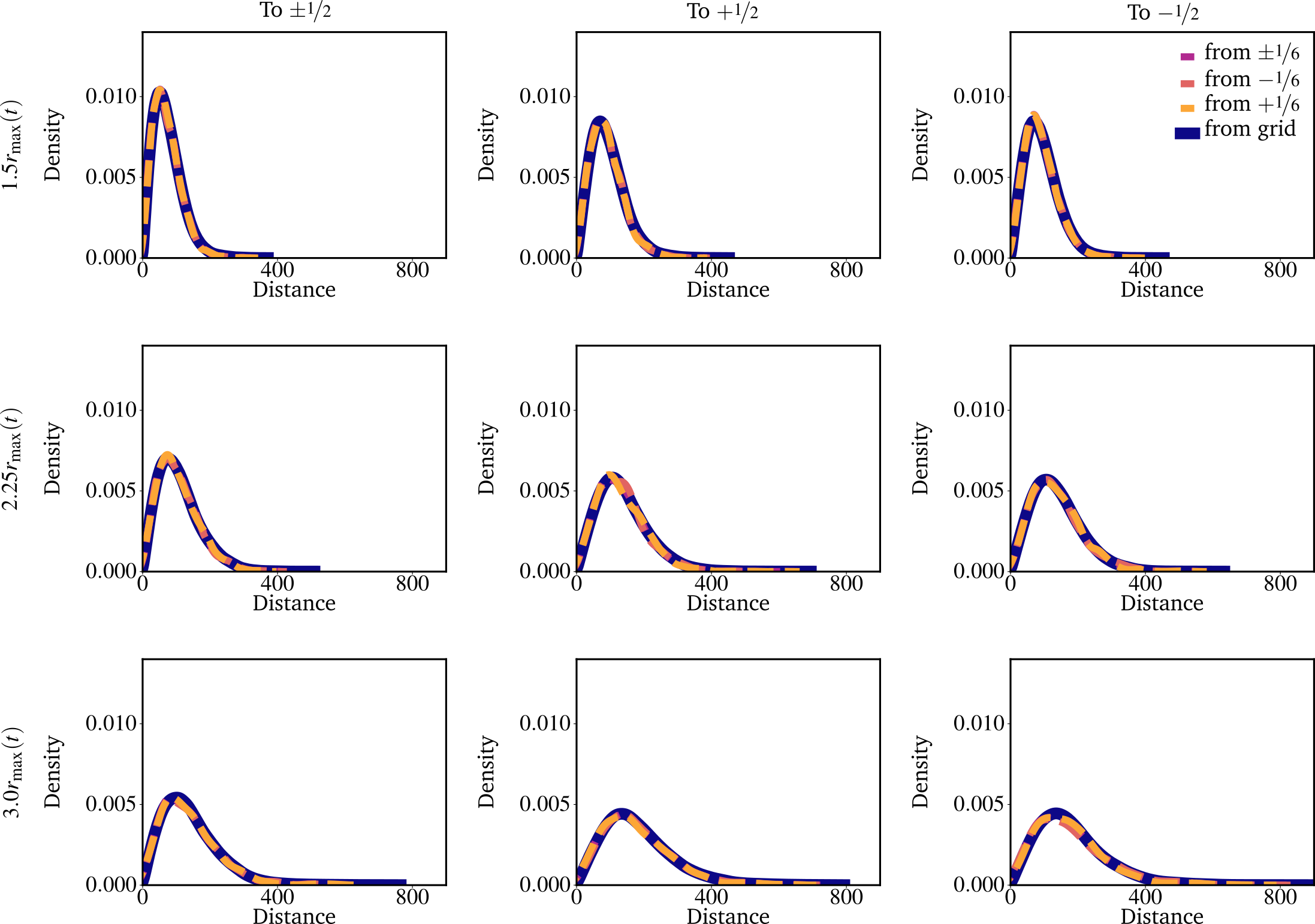}
    \caption{Kde plots of the density functions of the distance (measured in pixels) from $\pm \frac16$ ($-\frac16$, $+\frac16$) defects to the closest $\pm \frac12$ ($+\frac12$, $-\frac12$) defects and kde plots of the density functions of the distance from any gridpoint to the the closest $\pm \frac12$ ($+\frac12$, $-\frac12$) defects are shown for different coarse grain radii $\ravg$ (each row corresponds to one radius). The axis scaling is the same in all plots.}
    \label{fig:From_6_To_2}
\end{figure*}

A more central question in the current literature is whether orientational orders of different symmetry coexist in a correlated way, and whether their associated defects are spatially coupled. For this we first investigate the distance between a defect of order $p_1$ to the closest defect of order $p_2$, similar to the evaluations in \cite{Chiang_PNAS_2024}. For $p_1$ and $p_2$ we consider all possible combinations of $\pm \tfrac{1}{p_i}$ defects, only $+\tfrac{1}{p_i}$ defects, and only $-\tfrac{1}{p_i}$ defects, $i \in \{1,2\}$. 

We then compare the distribution of distances from a specific kind of defect of order $p_1$ to a specific kind of defect of order $p_2$ to the distribution of distances from any grid point to the same kind of defects of order $p_2$, using the Kolmogorov–Smirnov (KS) test. This test measures the maximum distance between the cumulative distribution functions of the two distributions. By construction, values close to $0$ indicate that the two distributions are nearly identical, whereas values close to $1$ reflect strong differences. If defects of order $p_1$ and $p_2$ were tightly coupled, we would expect large KS values.

\begin{figure*}[!hb]
    \centering
    \includegraphics[width=\linewidth]{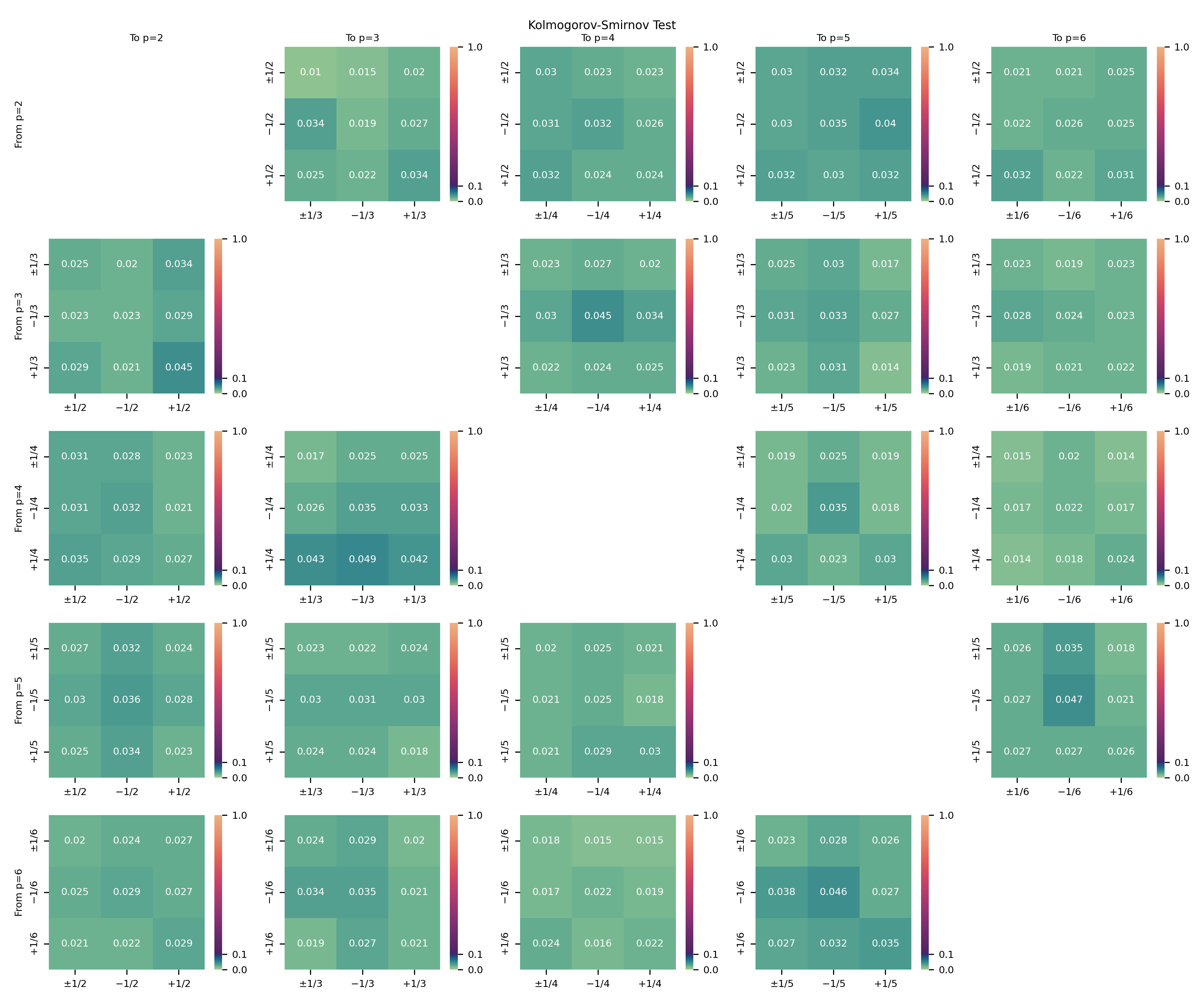}
    \caption{Results of the Kolmogorov-Smirnov test for all combinations of $\pm \frac{1}{p_1}$, $+\frac{1}{p_1}$, $-\frac{1}{p_1}$ defects with $\pm \frac{1}{p_2}$, $+\frac{1}{p_2}$, $-\frac{1}{p_2}$ defects for $\ravg=1.5r_{max}(t)$. We compare the distribution of distances from a specific kind of defects of order $p_1$ to a specific kind of defects of order $p_2$ to the distribution of distances from any grid point to the same kind of defects of order $p_2$. The result of the Kolmogorov–Smirnov is the maximum distance between the cumulative distribution functions of these two distributions. To easily distinguish between low and high values the background of the cells of the table are colored with a color scheme going from green to blue for values between $0.0$ and $0.1$ and with a color scheme from purple to orange for values between $0.1$ to $1.0$.}
    \label{fig:kstest_1_5}
\end{figure*}

\begin{figure*}[p]
    \centering
    \includegraphics[width=\linewidth]{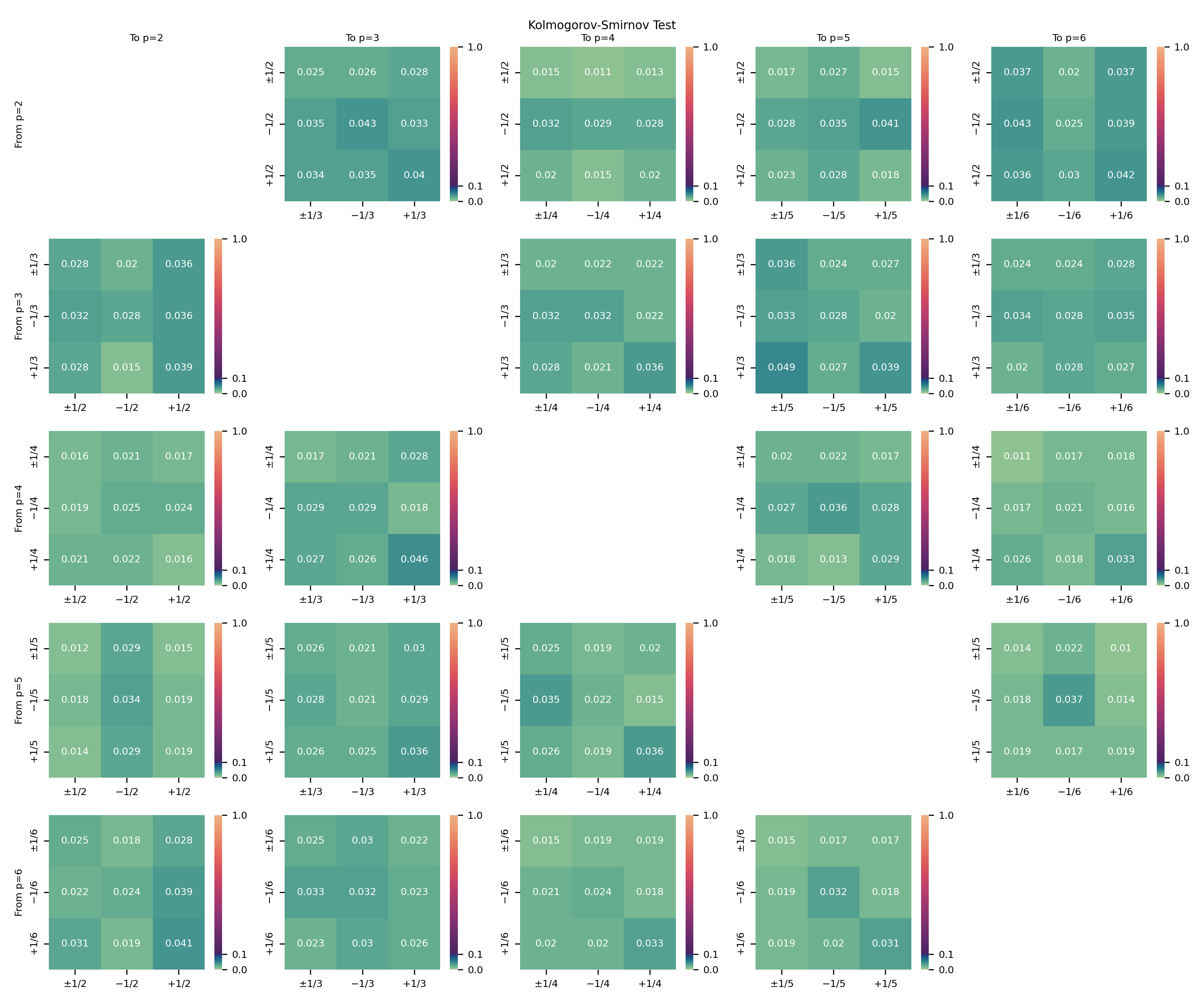}
    \caption{Results of the Kolmogorov-Smirnov test for all combinations of $\pm \frac{1}{p_1}$, $+\frac{1}{p_1}$, $-\frac{1}{p_1}$ defects with $\pm \frac{1}{p_2}$, $+\frac{1}{p_2}$, $-\frac{1}{p_2}$ defects for $\ravg=2.25r_{max}(t)$. We compare the distribution of distances from a specific kind of defects of order $p_1$ to a specific kind of defects of order $p_2$ to the distribution of distances from any grid point to the same kind of defects of order $p_2$. The result of the Kolmogorov–Smirnov is the maximum distance between the cumulative distribution functions of these two distributions. To easily distinguish between low and high values the background of the cells of the table are colored with a color scheme going from green to blue for values between $0.0$ and $0.1$ and with a color scheme from purple to orange for values between $0.1$ to $1.0$.}
    \label{fig:kstest_2_25}
\end{figure*}

\begin{figure*}[p]
    \centering
    \includegraphics[width=\linewidth]{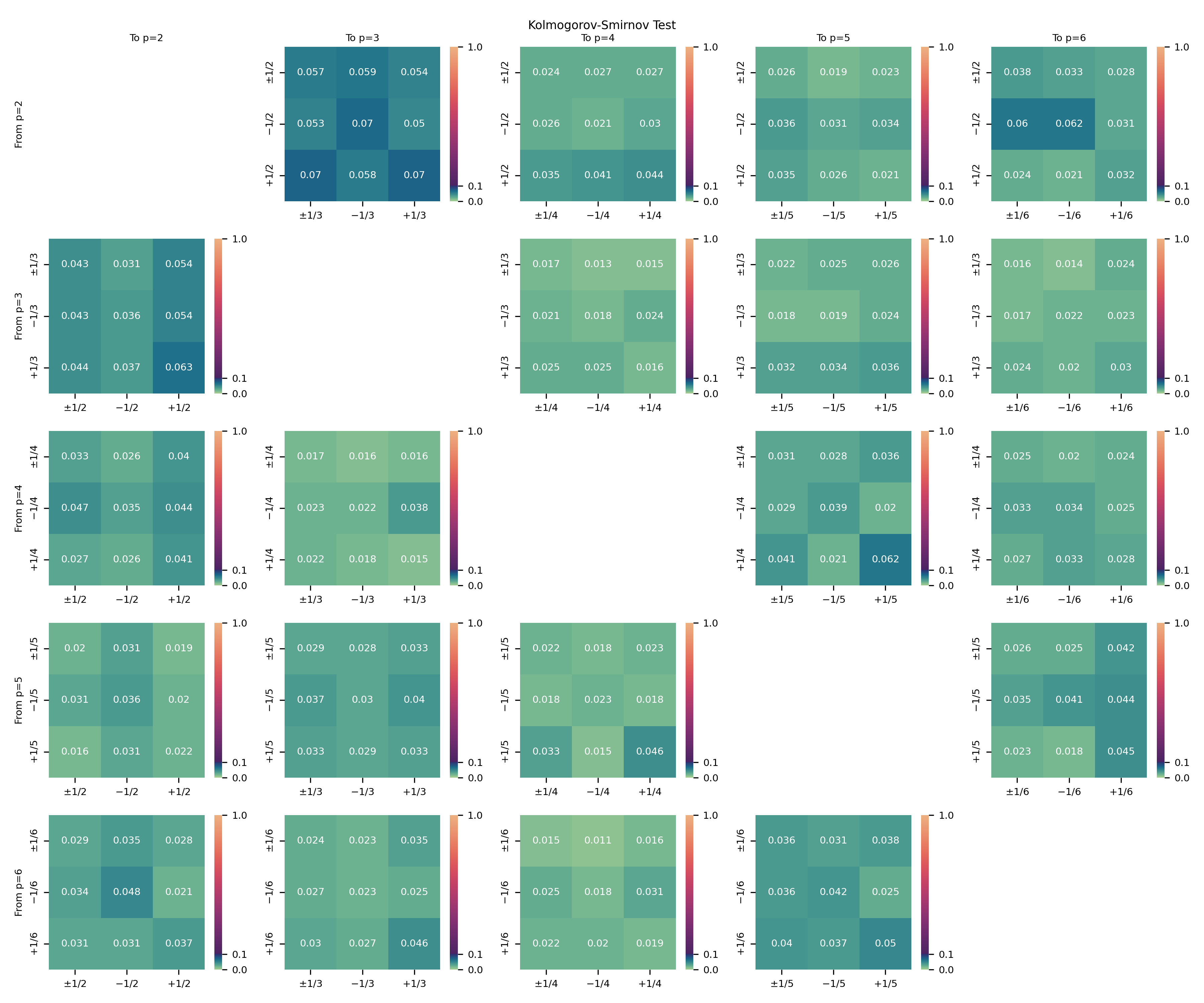}
    \caption{Results of the Kolmogorov-Smirnov test for all combinations of $\pm \frac{1}{p_1}$, $+\frac{1}{p_1}$, $-\frac{1}{p_1}$ defects with $\pm \frac{1}{p_2}$, $+\frac{1}{p_2}$, $-\frac{1}{p_2}$ defects for $\ravg=3.0r_{max}(t)$. We compare the distribution of distances from a specific kind of defects of order $p_1$ to a specific kind of defects of order $p_2$ to the distribution of distances from any grid point to the same kind of defects of order $p_2$. The result of the Kolmogorov–Smirnov is the maximum distance between the cumulative distribution functions of these two distributions. To easily distinguish between low and high values the background of the cells of the table are colored with a color scheme going from green to blue for values between $0.0$ and $0.1$ and with a color scheme from purple to orange for values between $0.1$ to $1.0$.}
    \label{fig:kstest_3_0}
\end{figure*}

We report the obtained KS values for all combinations in \nameref{S1_Appendix}: Figure \ref{fig:kstest_1_5} for $\ravg=1.5r_{max}(t)$, Figure \ref{fig:kstest_2_25} for $\ravg=2.25r_{max}(t)$, and Figure \ref{fig:kstest_3_0} for $\ravg=3.0r_{max}(t)$. To allow readers to easily distinguish between low and high values, the tables are color-coded, from green/blue for values $<0.1$ to purple/orange for values $>0.1$. Across all cases, the KS test yields values below $0.1$, indicating that the distributions are similar and that there is no strong positional correlation between distinct defect types. 

\begin{figure*}[p]
    \centering
    \includegraphics[width=\linewidth]{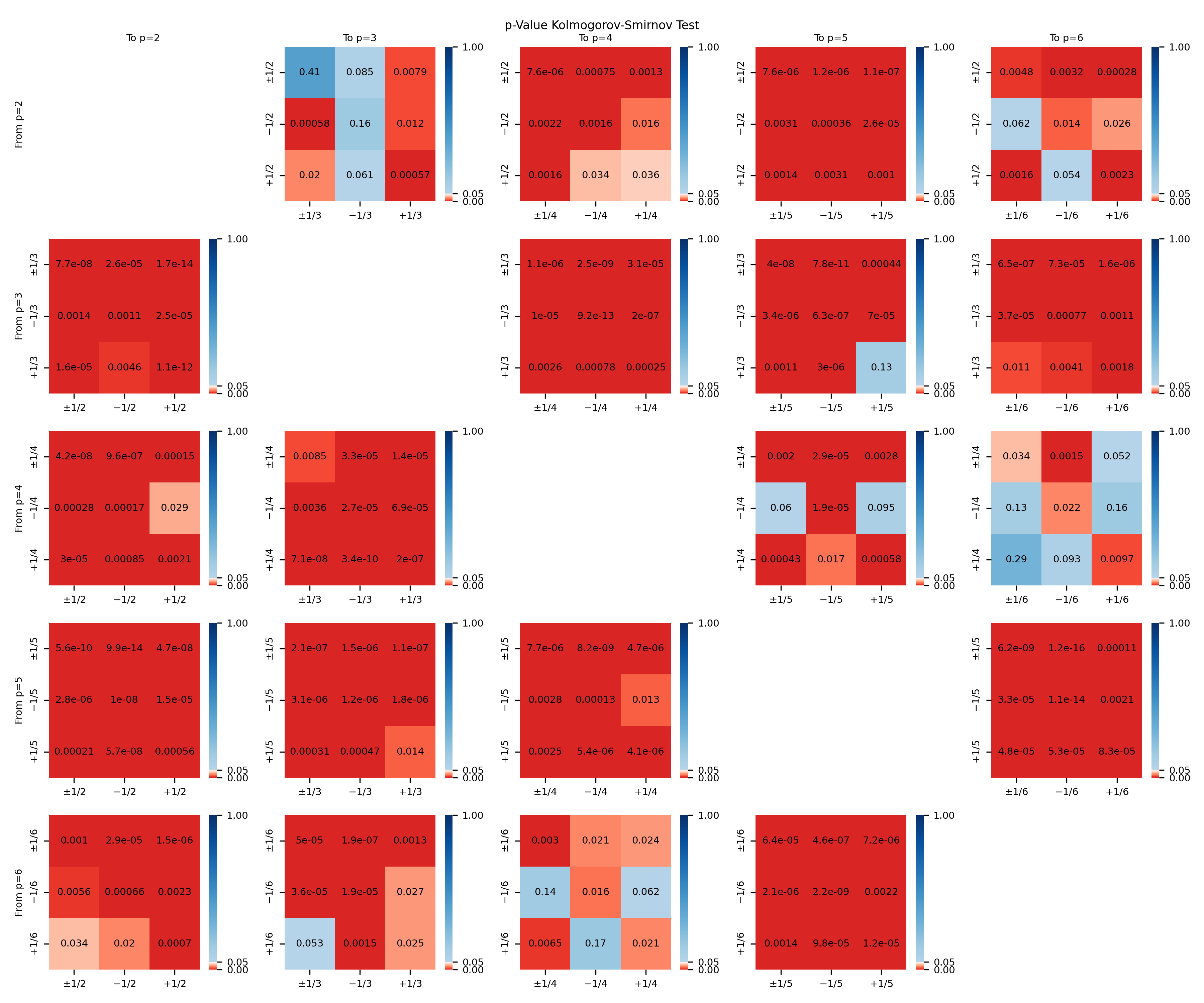}
    \caption{p-value of the Kolmogorov-Smirnov test  reported in Figure \ref{fig:kstest_1_5} (for defects obtained with a coarse-graining of $\ravg=1.5r_{max}(t)$). A low p-value $(\leq 0.05)$ means that the difference between the two cumulative distribution functions compared with the Kolmogorov-Smirnov test is significant and a high p-value means that it is not possible to show a difference between the two distributions. To guide the eye we colored the background of the cells in red for values $\leq 0.05$ and in blue for values $>0.05$.}
    \label{fig:kstest_pvalue_1_5}
\end{figure*}

\begin{figure*}[p]
    \centering
    \includegraphics[width=\linewidth]{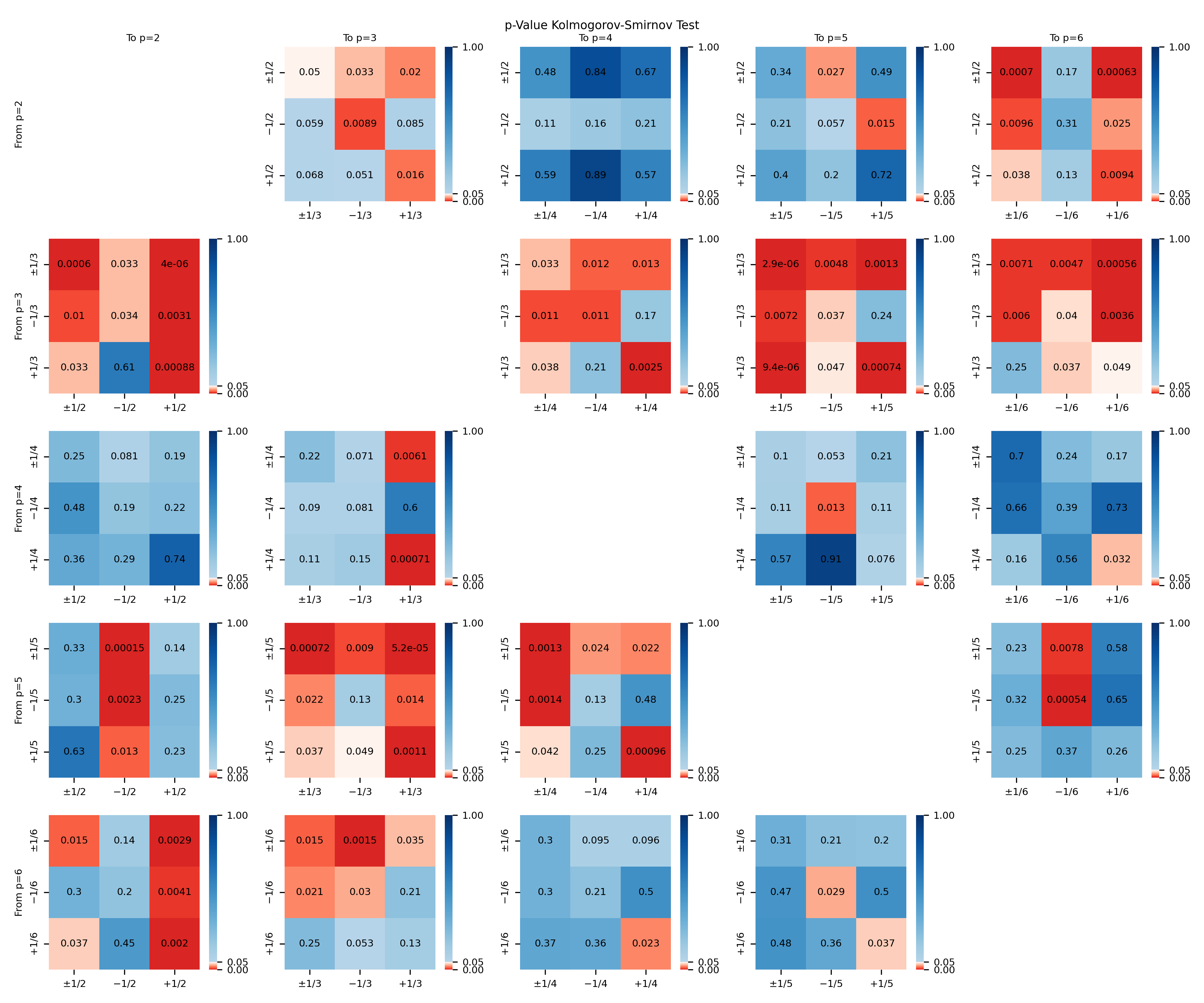}
    \caption{p-value of the Kolmogorov-Smirnov test  reported in Figure \ref{fig:kstest_2_25} (for defects obtained with a coarse-graining of $\ravg=2.25r_{max}(t)$). A low p-value $(\leq 0.05)$ means that the difference between the two cumulative distribution functions compared with the Kolmogorov-Smirnov test is significant and a high p-value means that it is not possible to show a difference between the two distributions. To guide the eye we colored the background of the cells in red for values $\leq 0.05$ and in blue for values $>0.05$.}
    \label{fig:kstest_pvalue_2_25}
\end{figure*}

\begin{figure*}[p]
    \centering
    \includegraphics[width=\linewidth]{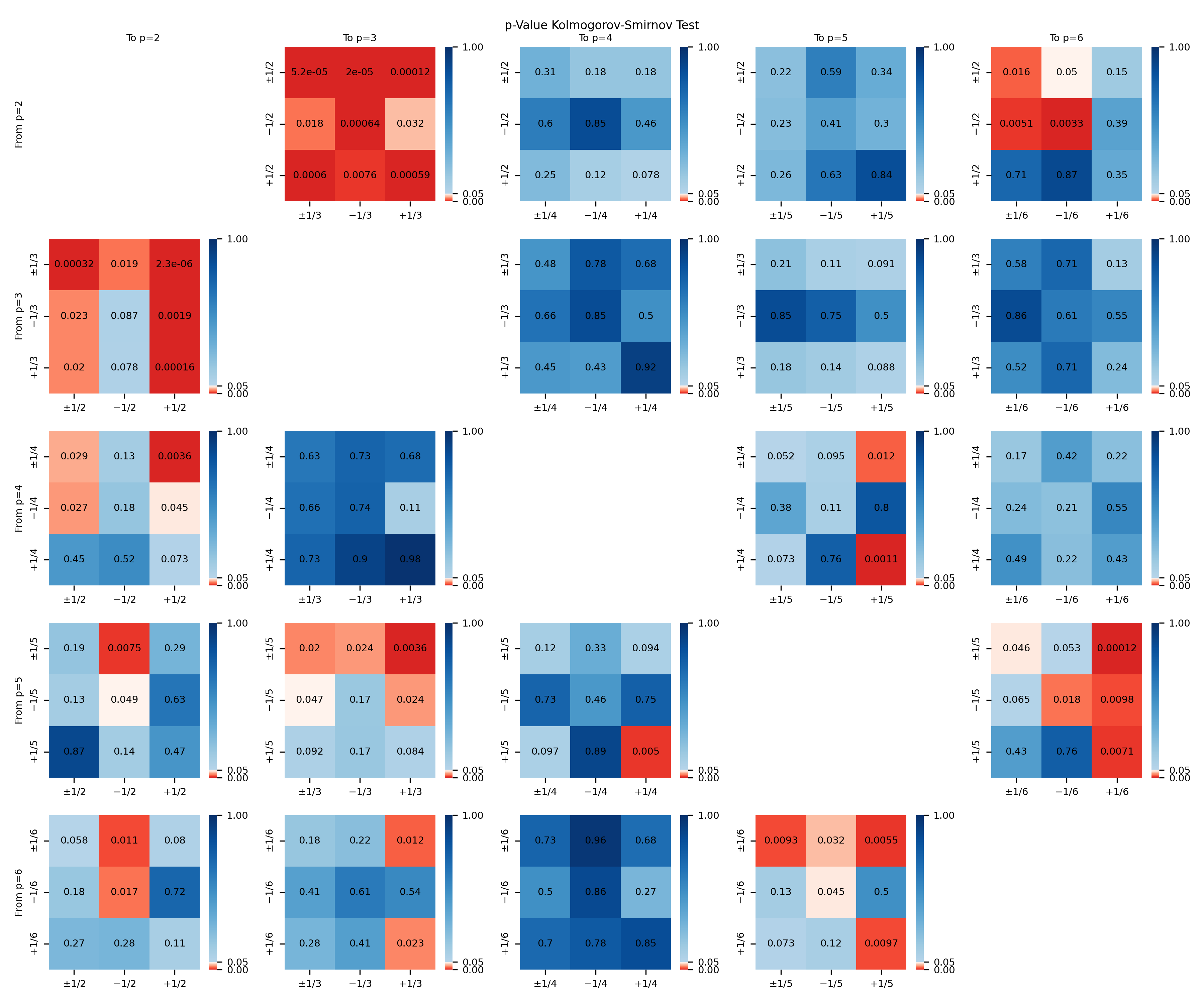}
    \caption{p-value of the Kolmogorov-Smirnov test  reported in Figure \ref{fig:kstest_3_0} (for defects obtained with a coarse-graining of $\ravg=3.0r_{max}(t)$). A low p-value $(\leq 0.05)$ means that the difference between the two cumulative distribution functions compared with the Kolmogorov-Smirnov test is significant and a high p-value means that it is not possible to show a difference between the two distributions. To guide the eye we colored the background of the cells in red for values $\leq 0.05$ and in blue for values $>0.05$.}
    \label{fig:kstest_pvalue_3_0}
\end{figure*}

To assess the statistical robustness of these comparisons we also examined the p-values of the KS test (Figures \ref{fig:kstest_pvalue_1_5}, \ref{fig:kstest_pvalue_2_25}, and \ref{fig:kstest_pvalue_3_0}). In this context p measures the probability of obtaining the observed results. A low p-value ($\leq 0.05$) indicates that the difference between distributions is significant, while higher values indicate that no statistically significant difference can be established. As expected, p-values tend to decrease with decreasing coarse-graining radius, since more defects provide larger sample sizes. However, these small differences are not our main focus: we are interested in whether correlations are both statistically significant and physically meaningful. 

To address this, we directly compared the density functions of the distances between nematic ($p=2$) and hexatic ($p=6$) defects, as this pair has been the subject of recent debate \cite{armengol2023epithelia,Chiang_PNAS_2024,Zhang_arxiv_2025}. As shown in Figures \ref{fig:From_2_To_6} and \ref{fig:From_6_To_2}, the distance distributions between nematic and hexatic defects are nearly indistinguishable from those obtained by random grid points. From this we conclude that, within our experimental resolution, there is no tight spatial correlation between $p=2$ and $p=6$ defects.
\begin{figure*}[htb]
    \centering
\includegraphics[width=0.9\linewidth]{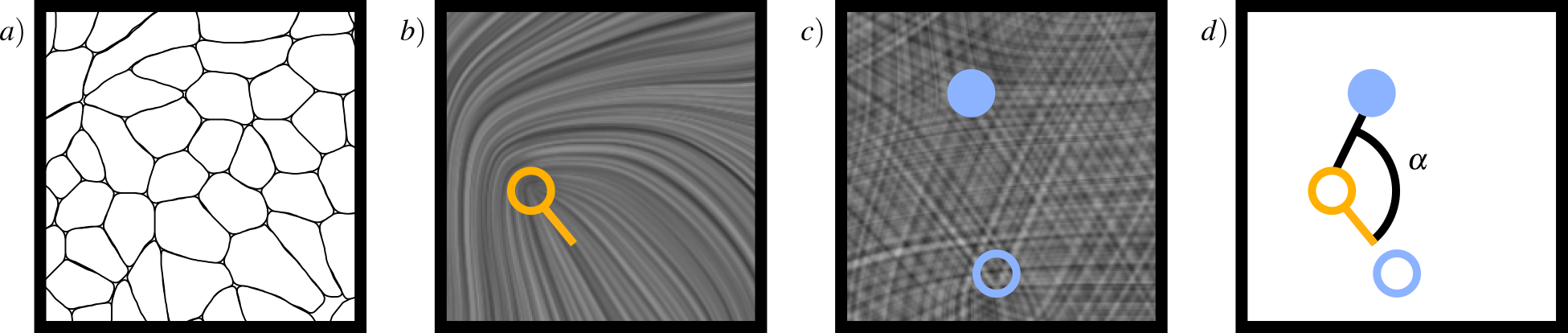}
    \caption{Illustration of the measured angle in Figure \ref{fig:DefectOrientation}. The cell outlines for the corresponding part of the experimental image are shown in $a)$, the LIC images with the defects for $p=2$ and $p=6$ are shown in $b)$ and $c)$, respectively. For the $+\frac12$ defect additionally the direction of the tail is marked. In $d)$ the defect positions for $p=2$ and $p=6$ are shown and the angle between the tail of the $+\frac12$ defect and the closest $-\frac16$ defect is marked with $\alpha$. As the closest $+\frac16$ defect is directly under the tail the angle between it and the tail of the $+\frac12$ defect would be $0^\circ$.}
    \label{fig:angle_defect}
\end{figure*}
\begin{figure*}[!htb]
    \centering
 \includegraphics[width=0.8\linewidth]{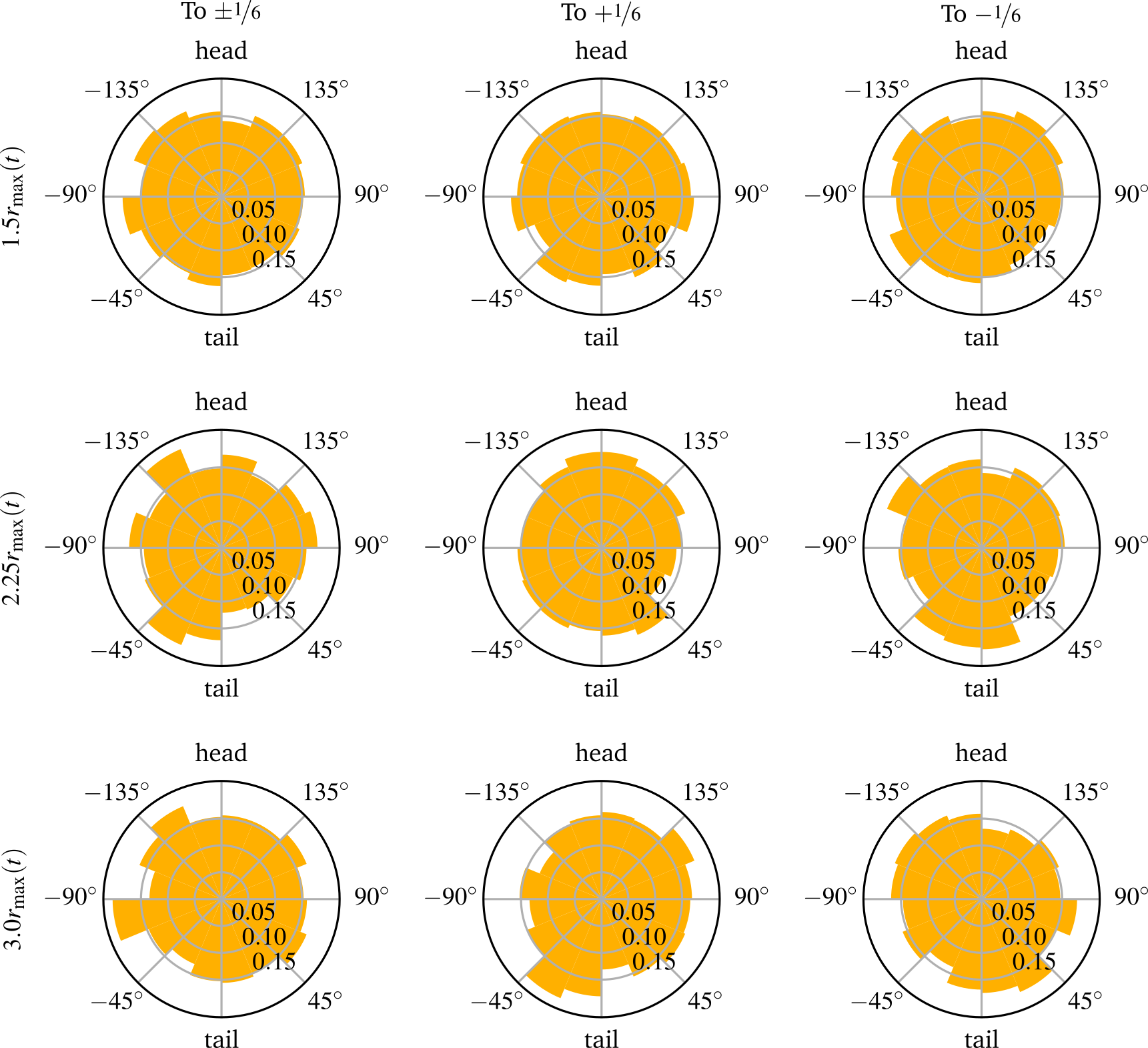}
    \caption{Density function of the angles between the orientation of a $+\frac12$ defect and the closest $\pm\frac16$ ($+\frac16$, $-\frac16$) defect. An angle of $0^\circ$ means that the closest $\pm \frac16$ defect lies at the tail of the $+\frac12$ defect, an angle of $180^\circ$ means that the closest $\pm \frac16$ ($+\frac16$, $-\frac16$) defect lies at the head of the $+\frac12$ defect. The orientation of the $+\frac12$ is calculated according to \cite{Vromans_SM_2016}. The axis scaling is the same in all plots.}
    \label{fig:DefectOrientation}
\end{figure*}

The distance-based analysis above probes positional correlations only, without considering relative orientations. To fully test whether different defect types interact in a directional manner, we next analyzed spatio–orientational correlations. We restrict this investigation to $+\tfrac{1}{2}$ defects because these are the only defects detected in our data that have a directionality. All other defects have some rotational symmetry, for example $-\tfrac12$ defects have a symmetry under a rotation of $120^\circ$ and are therefore not regarded here. 

We now evaluate the angle between the orientation of the $+\tfrac12$ defect and the closest $\pm\tfrac16$, $+\tfrac16$, or $-\tfrac16$ defect, and show the resulting density functions as polar plots in Figure \ref{fig:DefectOrientation}. The measured angle is illustrated in Figure \ref{fig:angle_defect}. By construction, a peak at $0^\circ$ would indicate that hexatic defects tend to localize at the tail of a $+\tfrac12$ defect, whereas a peak at $180^\circ$ would indicate alignment with the head.

Across all coarse-graining radii, evaluating these density functions for all found types of hexatic defects reveals that there is no spatio–orientational connection to the orientation of $+\tfrac12$ defects. This lack of alignment further supports the conclusion that nematic and hexatic defects in epithelial monolayers emerge largely independently, without systematic coupling at the level of defect orientation.
\section*{Discussion}

Independent of the coarse-graining radius $\ravg$, no connection between the locations of defects of different $p$-atic order could be found, and in particular we did not detect correlations between nematic and hexatic defects. While there are some studies on experimental data that include investigations of defect numbers \cite{armengol2023epithelia,eckert2023hexanematic}, studies focusing on defect positions are mainly restricted to computational data \cite{Chiang_PNAS_2024, Zhang_arxiv_2025}. A direct comparison of our results with these studies is not possible for two reasons: 
\begin{enumerate}
\item In both studies orientational defects are used for nematic order, but positional defects are used for hexatic order. While positional and orientational defects can be tightly related in cellular tissues, especially in a state close to the solid phase, they are not the same and we aim to keep them clearly separated. Positional defects are inherently tied to the cell scale, whereas orientational defects can be probed across multiple length scales depending on the averaging radius. By restricting our analysis to orientational defects, we access tissue-level symmetries rather than cell-level packing irregularities.
\item The studies\cite{Chiang_PNAS_2024,Zhang_arxiv_2025} mainly focus on the solid phase, which does not correspond to our data. However, in \cite{Zhang_arxiv_2025} it is reported that the correlation between hexatic and nematic defects weakens when going from the solid to the fluid phase. This trend is in agreement with what we see in our data, which correspond to the fluid-like state of epithelial monolayers. Our results therefore extend these observations into the biologically relevant regime where cells continuously divide, rearrange, and remodel their shapes.
\end{enumerate}

More broadly, our analysis demonstrates that constructing global $p$-atic fields and detecting defects therein provides a powerful bridge between experimental data or agent-based models of cells and coarse-grained continuum descriptions. While first models including higher-order $p$-atics have been proposed \cite{giomi2022hydrodynamic,giomi2022long} and applied to epithelial tissues \cite{krommydas2023hydrodynamic,armengol2023epithelia}, it remains unclear which $p$-atic orders are most relevant. As we did not find positional correlations between defects of distinct $p$, we propose that coarse-grained liquid crystal frameworks for epithelial tissue may need to integrate multiple $p$ simultaneously. Following the same reasoning, the exclusive focus on the relation between $p=2$ and $p=6$ should be reconsidered. Indeed, our results (Fig.~\ref{fig:defects_over_time}) suggest an intriguing even–odd effect in defect numbers, motivating further study. There is also no indication that $p=4$ is less important: correlations between tetratic order and cell division have recently been reported \cite{Cislo_NaturePhysics_2023}, and it will be an interesting future direction to investigate whether tetratic defects are directly linked to cell division events.

Finally, the choice of length scale is crucial, both for constructing coarse-grained fields and for interpreting defect statistics. It remains unclear whether global fields and defects should be calculated closer to the cell scale or the tissue scale, and whether the same scale should be used for all defect types. Our findings suggest no clear dominance of one defect type at a given scale. However, mathematical tests in isolation cannot fully capture the underlying biological complexity. Connecting different defects to functional cellular or tissue-level processes—similar to how $+\tfrac{1}{2}$ defects were linked to cell extrusion \cite{saw2017topological}—could provide decisive insights. In this sense, the presence and dynamics of $p$-atic defects may serve not only as order-parameter singularities but also as markers of key biological events in morphogenesis.

\section*{Conclusion and Outlook}  
Our study highlights how irregular cell shapes in epithelial monolayers naturally generate a spectrum of $p$-atic orders and their associated topological defects. By bridging cell-resolved data with continuum concepts, we show that multiple symmetries can coexist without strong correlations, suggesting that tissues are not restricted to a single symmetry class. This opens several exciting directions: investigating how higher-order defects couple to biological processes such as division, extrusion, or migration; testing whether even–odd asymmetries in defect statistics hold across different epithelial systems; and extending coarse-grained models to explicitly integrate multiple $p$ simultaneously. More broadly, our approach demonstrates how concepts from soft condensed matter physics—symmetry, topology, and scale—can be leveraged to uncover hidden organizing principles in living matter.
\section*{Author contributions}
L.H.: Investigation (equal), Methodology (equal), Software (equal), Visualization (equal), Writing – original draft, Writing – review \& editing (equal). 
G.O.: Investigation (equal), Methodology (equal), Software (equal), Visualization (equal), A.R., G.R.R.: Investigation (support), Software (support)
V.G.: Experimental setup (lead), Segmentation (lead).
A.D.: Methodology (support), Funding acquisition, Writing – review \& editing (equal).
A.V.: Conceptualization, Investigation (support), Methodology (support), Supervision, Funding acquisition, Writing – review \& editing (equal).

\section*{Conflicts of interest}
There are no conflicts to declare.

\section*{Data availability}
For all tasks we provide the considered Python code, see \cite{zenodo}.

\section*{Acknowledgments}
We acknowledge fruitful discussions with Björn Böttcher, Brendan Tobin and Emma Happel. AD acknowledges funding from the Novo Nordisk Foundation (grant No. NNF18SA0035142 and NERD grant No. NNF21OC0068687), Villum Fonden (grant No. 29476), and the European Union (ERC, PhysCoMeT, 101041418). AV acknowledges funding from the German Research Foundation (Award FOR3013 "Vector- and tensor-valued surface PDEs") and computing resources provided by JSC through MORPH and by ZIH through WIR. 

\section*{Supporting information}
\label{S1_Appendix}
\phantom{mylong text}
\begin{figure*}[b]
\vspace{2em}
    \centering
\includegraphics[width=0.83\linewidth]{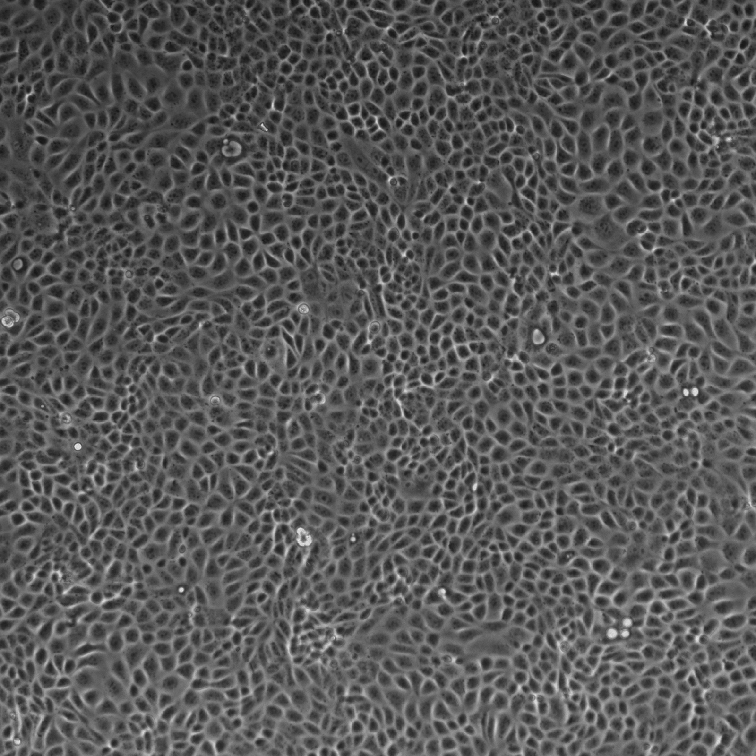}
    \caption{Raw experimental image from Frame 25, which was used in Figure \ref{fig:scatter_plot}.}
    \label{fig:frame_25}
\end{figure*}
\phantom{my long text}

\begin{figure*}[p]
    \centering
    \includegraphics[width=0.83\linewidth]{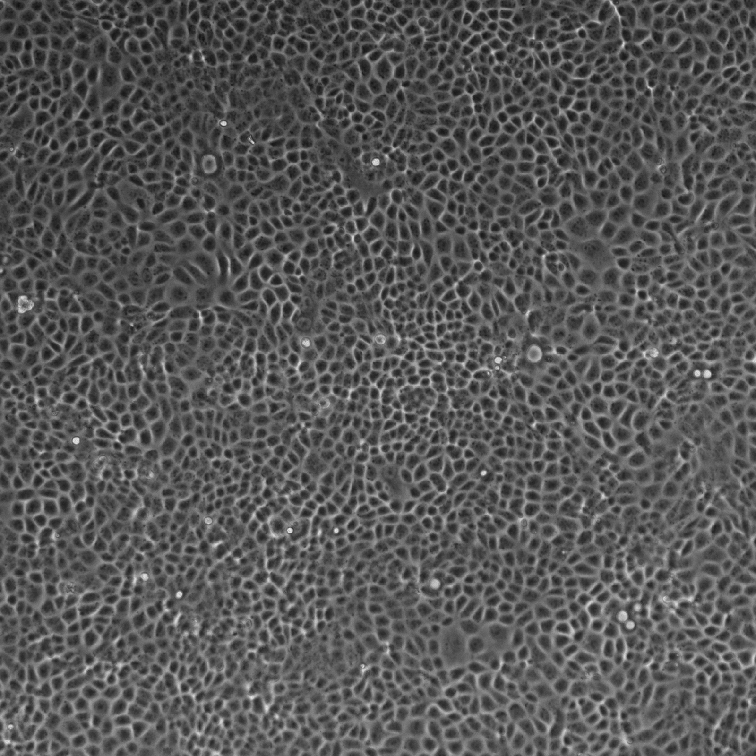}
    \caption{Raw experimental image from Frame 50, which was used in Figure \ref{fig:scatter_plot}.}
    \label{fig:frame_50}
\end{figure*}


\renewcommand\refname{References}

\bibliography{bib} 
\bibliographystyle{abbrv} 

\end{document}

%% file: macros_Defects.tex
\newcommand{\QTensorCompA}{Q_0}
\newcommand{\QTensorCompB}{Q_1}
\newcommand{\QTensorCompAInt}{\QTensorCompA^{int}}
\newcommand{\QTensorCompBInt}{\QTensorCompB^{int}}
\newcommand{\ravg}{r_{avg}}

%% file: arxiv.bbl
\begin{thebibliography}{10}

\bibitem{alert2022active}
R.~Alert, J.~Casademunt, and J.-F. Joanny.
\newblock Active turbulence.
\newblock {\em Annual Review of Condensed Matter Physics}, 13(1):143--170, 2022.

\bibitem{alert2020physical}
R.~Alert and X.~Trepat.
\newblock Physical models of collective cell migration.
\newblock {\em Annual Review of Condensed Matter Physics}, 11(1):77--101, 2020.

\bibitem{armengol2023epithelia}
J.-M. Armengol-Collado, L.~N. Carenza, J.~Eckert, D.~Krommydas, and L.~Giomi.
\newblock Epithelia are multiscale active liquid crystals.
\newblock {\em Nature Physics}, 19:1773–1779, 2023.

\bibitem{armengol2024hydrodynamics}
J.-M. Armengol-Collado, L.~N. Carenza, and L.~Giomi.
\newblock Hydrodynamics and multiscale order in confluent epithelia.
\newblock {\em eLife}, 13:e86400, 2024.

\bibitem{PhysRevLett.107.155704}
E.~P. Bernard and W.~Krauth.
\newblock Two-step melting in two dimensions: first-order liquid-hexatic transition.
\newblock {\em Physical Review Letters}, 107:155704, 2011.

\bibitem{PhysRevLett.74.2519}
P.~Bladon and D.~Frenkel.
\newblock Dislocation unbinding in dense two-dimensional crystals.
\newblock {\em Physical Review Letters}, 74:2519--2522, 1995.

\bibitem{Bowick_2017}
M.~J. Bowick, O.~V. Manyuhina, and F.~Serafin.
\newblock Shapes and singularities in triatic liquid-crystal vesicles.
\newblock {\em Europhysics Letters}, 117(2):26001, 2017.

\bibitem{Chiang_PNAS_2024}
M.~Chiang, A.~Hopkins, B.~Loewe, M.~C. Marchetti, and D.~Marenduzzo.
\newblock Intercellular friction and motility drive orientational order in cell monolayers.
\newblock {\em Proceedings of the National Academy of Sciences}, 121(40):e2319310121, 2024.

\bibitem{Cislo_NaturePhysics_2023}
D.~Cislo, F.~Yang, H.~Qin, A.~Pavlopoulos, M.~Bowick, and S.~Streichan.
\newblock Active cell divisions generate fourfold orientationally ordered phase in living tissue.
\newblock {\em Nature Physics}, 19(8):1201–1210, 2023.

\bibitem{de_Gennes_book}
P.~G. de~Gennes and J.~Prost.
\newblock {\em The physics of liquid crystals}.
\newblock Second Edition, Clarendon Press, Oxford, 1993.

\bibitem{duclos2017topological}
G.~Duclos, C.~Erlenk{\"a}mper, J.-F. Joanny, and P.~Silberzan.
\newblock Topological defects in confined populations of spindle-shaped cells.
\newblock {\em Nature Physics}, 13(1):58--62, 2017.

\bibitem{durand2019thermally}
M.~Durand and J.~Heu.
\newblock Thermally driven order-disorder transition in two-dimensional soft cellular systems.
\newblock {\em Physical Review Letters}, 123(18):188001, 2019.

\bibitem{eckert2023hexanematic}
J.~Eckert, B.~Ladoux, R.-M. M{\`e}ge, L.~Giomi, and T.~Schmidt.
\newblock Hexanematic crossover in epithelial monolayers depends on cell adhesion and cell density.
\newblock {\em Nature Communications}, 14(1):5762, 2023.

\bibitem{cphc.200900755}
U.~Gasser, C.~Eisenmann, G.~Maret, and P.~Keim.
\newblock Melting of crystals in two dimensions.
\newblock {\em ChemPhysChem.}, 11(5):963--970, 2010.

\bibitem{giomi2022hydrodynamic}
L.~Giomi, J.~Toner, and N.~Sarkar.
\newblock Hydrodynamic theory of p-atic liquid crystals.
\newblock {\em Physical Review E}, 106(2):024701, 2022.

\bibitem{giomi2022long}
L.~Giomi, J.~Toner, and N.~Sarkar.
\newblock Long-ranged order and flow alignment in sheared p-atic liquid crystals.
\newblock {\em Physical Review Letters}, 129(6):067801, 2022.

\bibitem{PhysRevLett.41.121}
B.~I. Halperin and D.~R. Nelson.
\newblock Theory of two-dimensional melting.
\newblock {\em Physical Review Letters}, 41:121--124, 1978.

\bibitem{Happel2025.01.03.631196}
L.~Happel, G.~Oberschelp, V.~Grudtsyna, H.~P. Jain, R.~Sknepnek, A.~Doostmohammadi, and A.~Voigt.
\newblock Quantifying the shape of cells - from minkowski tensors to p-atic order.
\newblock {\em bioRxiv}, 2025.

\bibitem{kawaguchi2017topological}
K.~Kawaguchi, R.~Kageyama, and M.~Sano.
\newblock Topological defects control collective dynamics in neural progenitor cell cultures.
\newblock {\em Nature}, 545(7654):327--331, 2017.

\bibitem{krommydas2023hydrodynamic}
D.~Krommydas, L.~N. Carenza, and L.~Giomi.
\newblock Hydrodynamic enhancement of p-atic defect dynamics.
\newblock {\em Physical Review Letters}, 130(9):098101, 2023.

\bibitem{Laine_2019}
R.~F. Laine, K.~L. Tosheva, N.~Gustafsson, R.~D.~M. Gray, P.~Almada, D.~Albrecht, G.~T. Risa, F.~Hurtig, A.-C. Lindås, B.~Baum, J.~Mercer, C.~Leterrier, P.~M. Pereira, S.~Culley, and R.~Henriques.
\newblock Nanoj: a high-performance open-source super-resolution microscopy toolbox.
\newblock {\em Journal of Physics D: Applied Physics}, 52(16):163001, 2019.

\bibitem{Laramee_IEEE_2003}
R.~Laramee, B.~Jobard, and H.~Hauser.
\newblock Image space based visualization of unsteady flow on surfaces.
\newblock In {\em IEEE Visualization, 2003. VIS 2003.}, pages 131--138, 2003.

\bibitem{li2018role}
Y.-W. Li and M.~P. Ciamarra.
\newblock Role of cell deformability in the two-dimensional melting of biological tissues.
\newblock {\em Physical Review Materials}, 2(4):045602, 2018.

\bibitem{Loewe_PRL_2020}
B.~Loewe, M.~Chiang, D.~Marenduzzo, and M.~Marchetti.
\newblock Solid-liquid transition of deformable and overlapping active particles.
\newblock {\em Physical Review Letters}, 125:038003, 2020.

\bibitem{maroudas2021topological}
Y.~Maroudas-Sacks, L.~Garion, L.~Shani-Zerbib, A.~Livshits, E.~Braun, and K.~Keren.
\newblock Topological defects in the nematic order of actin fibres as organization centres of hydra morphogenesis.
\newblock {\em Nature Physics}, 17(2):251--259, 2021.

\bibitem{monfared2023mechanical}
S.~Monfared, G.~Ravichandran, J.~Andrade, and A.~Doostmohammadi.
\newblock Mechanical basis and topological routes to cell elimination.
\newblock {\em eLife}, 12:e82435, 2023.

\bibitem{PhysRevLett.58.1200}
C.~A. Murray and D.~H. Van~Winkle.
\newblock Experimental observation of two-stage melting in a classical two-dimensional screened coulomb system.
\newblock {\em Physical Review Letters}, 58:1200--1203, 1987.

\bibitem{Nelson_JP_1987}
D.~Nelson and L.~Peliti.
\newblock Fluctuations in membranes with crystalline and hexatic order.
\newblock {\em Journal de physique}, 48(7):1085--1092, 1987.

\bibitem{PhysRevB.19.2457}
D.~R. Nelson and B.~I. Halperin.
\newblock Dislocation-mediated melting in two dimensions.
\newblock {\em Physical Review B}, 19:2457--2484, 1979.

\bibitem{zenodo}
G.~Oberschelp, A.~Richter, G.~Rode, and L.~Happel.
\newblock {\em Upload on Zenodo of the our python code- zenodo doi will be added after the Review}.

\bibitem{j.1749-6632.1949.tb27296.x}
L.~Onsager.
\newblock The effects of shape on the interaction of colloidal particales.
\newblock {\em Annals of the New York Academy of Sciences}, 51(4):627--659, 1949.

\bibitem{10.1145/1276377.1276446}
J.~Palacios and E.~Zhang.
\newblock Rotational symmetry field design on surfaces.
\newblock {\em ACM Trans. Graph.}, 26(3):55–es, 2007.

\bibitem{Palacios_IEEE_2011}
J.~Palacios and E.~Zhang.
\newblock Interactive visualization of rotational symmetry fields on surfaces.
\newblock {\em IEEE Transactions on Visualization and Computer Graphics}, 17(7):947--955, 2011.

\bibitem{pasupalak2020hexatic}
A.~Pasupalak, L.~Yan-Wei, R.~Ni, and M.~P. Ciamarra.
\newblock Hexatic phase in a model of active biological tissues.
\newblock {\em Soft matter}, 16(16):3914--3920, 2020.

\bibitem{Pylv504744}
J.~W. Pylv{\"a}n{\"a}inen, R.~F. Laine, B.~M. Saraiva, S.~Ghimire, G.~Follain, R.~Henriques, and G.~Jacquemet.
\newblock Fast4dreg: Fast registration of 4d microscopy datasets.
\newblock {\em bioRxiv}, 2022.

\bibitem{10.1145/1356682.1356683}
N.~Ray, B.~Vallet, W.~C. Li, and B.~L\'{e}vy.
\newblock N-symmetry direction field design.
\newblock {\em ACM Trans. Graph.}, 27(2):Article 10, 2008.

\bibitem{saw2017topological}
T.~B. Saw, A.~Doostmohammadi, V.~Nier, L.~Kocgozlu, S.~Thampi, Y.~Toyama, P.~Marcq, C.~T. Lim, J.~M. Yeomans, and B.~Ladoux.
\newblock Topological defects in epithelia govern cell death and extrusion.
\newblock {\em Nature}, 544(7649):212--216, 2017.

\bibitem{vtkBook}
W.~Schroeder, K.~Martin, and B.~Lorensen.
\newblock {\em The Visualization Toolkit (4th ed.)}.
\newblock Kitware, 2006.

\bibitem{skogvoll2023unified}
V.~Skogvoll, J.~R{\o}nning, M.~Salvalaglio, and L.~Angheluta.
\newblock A unified field theory of topological defects and non-linear local excitations.
\newblock {\em npj Computational Materials}, 9(1):122, 2023.

\bibitem{cellpose2021}
C.~Stringer, T.~Wang, M.~Michaelos, and M.~Pachitariu.
\newblock Cellpose: a generalist algorithm for cellular segmentation.
\newblock {\em Nature Methods}, 18:100–106, 2021.

\bibitem{trackmate2017}
J.-Y. Tinevez, N.~Perry, J.~Schindelin, G.~M. Hoopes, G.~D. Reynolds, E.~Laplantine, S.~Y. Bednarek, S.~L. Shorte, and K.~W. Eliceiri.
\newblock Trackmate: An open and extensible platform for single-particle tracking.
\newblock {\em Methods}, 115:80--90, 2017.

\bibitem{scikit-image}
S.~van~der Walt, J.~L. {S}ch\"onberger, J.~{Nunez-Iglesias}, F.~{B}oulogne, J.~D. {W}arner, N.~{Y}ager, E.~{G}ouillart, T.~{Y}u, and the scikit-image contributors.
\newblock scikit-image: image processing in {P}ython.
\newblock {\em PeerJ}, 2:e453, 6 2014.

\bibitem{Vromans_SM_2016}
A.~J. Vromans and L.~Giomi.
\newblock Orientational properties of nematic disclinations.
\newblock {\em Soft Matter}, 12:6490--6495, 2016.

\bibitem{wang2018brownian}
P.-Y. Wang and T.~G. Mason.
\newblock A brownian quasi-crystal of pre-assembled colloidal penrose tiles.
\newblock {\em Nature}, 561(7721):94--99, 2018.

\bibitem{wenzel2021defects}
D.~Wenzel, M.~Nestler, S.~Reuther, M.~Simon, and A.~Voigt.
\newblock Defects in active nematics--algorithms for identification and tracking.
\newblock {\em Computational Methods in Applied Mathematics}, 21(3):683--692, 2021.

\bibitem{Wenzel_PRE_2021}
D.~Wenzel and A.~Voigt.
\newblock Multiphase field models for collective cell migration.
\newblock {\em Physical Review E}, 184:054410, 2021.

\bibitem{Wojciechowski}
K.~W. Wojciechowski and D.~Frenkel.
\newblock Tetratic phase in the planar hard square system?
\newblock {\em Computational Methods in Science and Technology}, 10:235--255, 2004.

\bibitem{YU2023}
T.~Yu and T.~G. Mason.
\newblock Heptatic liquid quasi-crystals by colloidal lithographic pre-assembly.
\newblock {\em Journal of Colloid and Interface Science}, 2023.

\bibitem{PhysRevLett.82.2721}
K.~Zahn, R.~Lenke, and G.~Maret.
\newblock Two-stage melting of paramagnetic colloidal crystals in two dimensions.
\newblock {\em Physical Review Letters}, 82:2721--2724, 1999.

\bibitem{Zhang_arxiv_2025}
J.~Zhang, C.~W. Chan, B.~Li, and R.~Zhang.
\newblock Lattice and orientational defects mediate collective transport of confluent cells.
\newblock {\em arXiv preprint 2506.04068}, 2025.

\end{thebibliography}
